\documentclass[A4,11pt]{amsart}
\usepackage{{amsmath,amsfonts,latexsym, amstext, amssymb}}
\usepackage{graphicx,epstopdf}
\pdfoutput=1

\newtheorem{definition}{Definition}[section]

\newtheorem{thm}[definition]{Theorem}

\newtheorem{lemma}[definition]{Lemma}

\newcommand{\CG}{\hbox{{$\mathcal G$}}}

\newcommand{\CP}{\hbox{{$\mathcal P$}}}
\renewcommand{\P}{{\CP}}

\newcommand{\CX}{\hbox{{$\mathcal X$}}}

\newcommand{\g}{\mathrm g}
\newcommand{\m}{m}
\newcommand{\n}{k}

\newcommand{\C}{\mathbb{C}}
\newcommand{\R}{\mathbb{R}}

\newcommand{\Z}{\mathbb{Z}}
\newcommand{\N}{\mathbb{N}}

\newcommand{\del}{\partial}

\newcommand{\extd}{{\rm d}}
\newcommand{\trace}{{\rm Tr}}
\newcommand{\isom}{{\cong}}
\newcommand{\eps}{{\epsilon}}
\newcommand{\tens}{\mathop{\otimes}}
\newcommand{\la}{{\triangleright}}
\newcommand{\ra}{{\triangleleft}}

\newcommand{\Ad}{{\rm Ad}}
\newcommand{\ad}{{\rm ad}}

\newcommand{\<}{\langle}
\renewcommand{\>}{\rangle}

\def\rcross{{\triangleright\!\!\!<}}
\def\lcross{{>\!\!\!\triangleleft}}
\def\rcocross{{\blacktriangleright\!\!<}}

\def\rlbicross{{\triangleright\!\!\!\blacktriangleleft}}
\def\lrbicross{{\blacktriangleright\!\!\!\triangleleft}}
\def\dcross{{\bowtie}}
\def\codcross{{\blacktriangleright\!\!\blacktriangleleft}}
\def\rbiprod{{\cdot\kern-.33em\triangleright\!\!\!<}}
\def\lbiprod{{>\!\!\!\triangleleft\kern-.33em\cdot}}

\renewcommand{\o}{{}_{\scriptscriptstyle(1)}}
\renewcommand{\t}{{}_{\scriptscriptstyle(2)}}
\newcommand{\thr}{{}_{\scriptscriptstyle(3)}}

\newcommand{\bz}{{}_{{\scriptscriptstyle(0)}}}

\newcommand{\bo}{{}_{{\scriptscriptstyle(1)}}}

\def\Ad{\mbox{Ad}}
\def\ad{\mbox{ad}}
\def\cd{\!\cdot\!}

\def\bpm{\begin{pmatrix}}
\def\epm{\end{pmatrix}}

\newcommand{\gothg}{\mathfrak g }

\newcommand{\RR}{\mathbb{R}}
\newcommand{\CC}{\mathbb{C}}
\newcommand{\tr}{{\rm tr}}
\def\bea{\begin{eqnarray}}
\def\eea{\end{eqnarray}}

\begin{document}
\begin{flushright}
EMPG-08-09\\
\end{flushright}
%
%
%
%
%
%
%
%

\title[$q$-deformation and semidualisation in 3d quantum gravity]
{$q$-deformation and semidualisation \\ in 3d quantum gravity}

\author{S.~Majid}
\address{School of Mathematical Sciences\\
Queen Mary, University of London\\ 327 Mile End Rd,  London E1
4NS, UK\\ E-mail {\tt s.majid@qmul.ac.uk}}
\author{B.~J.~Schroers}
\address{Department of Mathematics and Maxwell Institute for Mathematical Sciences \\ Heriot-Watt University \\
Edinburgh EH14 4AS, UK \\ E-mail {\tt bernd@ma.hw.ac.uk}}

\subjclass{58B32, 58B34, 20C05} \keywords{Quantum gravity, quantum groups, cosmological constant, bicrossproduct}

\date{June 2008. Revised January 2009}

\maketitle

\begin{abstract} We explore in detail the role in euclidean 3d quantum gravity of quantum Born reciprocity or `semidualisation'. The latter is an algebraic operation defined using quantum group methods that interchanges position and momentum. Using this we are able to clarify the structural relationships between the effective noncommutative geometries that have been discussed in the 
context of 3d gravity. We show that the spin model based on $D(U(su_2))$ for quantum gravity without cosmological constant is the semidual of a quantum  particle on a three-sphere,  while  the bicrossproduct (DSR) model based on $\C[\R^2\lcross\R]\lrbicross U(su_2)$ is the semidual of a quantum  particle 
on hyperbolic space. We show further how the different models are all specific limits of  $q$-deformed models with $q=e^{-\hbar   \sqrt{-\Lambda}/m_p}$ where $m_p$ is the Planck mass  and $\Lambda $ is the cosmological  constant, and argue that semidualisation   interchanges $m_p\leftrightarrow l_c$, 
  where $l_c$ is the cosmological length scale $l_c=1/\sqrt{|\Lambda|}$.    We investigate the  physics of  semidualisation by studying  representation theory. In both the spin model and its semidual we show that irreducible representations have a physical picture as solutions of a respectively noncommutative/curved wave equation. We explain, moreover, that the $q$-deformed model, at a certain algebraic level, is self-dual under semidualisation.  \end{abstract}

\bibliographystyle{plain}

\section{Introduction}

Whatever quantum gravity actually is, it must provide classical continuum geometry at macroscopic scales and involve corrections at  the Planck scale. In recent years it has become more widely accepted that these corrections should, at least at first order, be described by some kind of noncommutative geometry in which coordinate algebras are noncommutative or `quantum'. A useful setting for exploring this idea is provided by 3d quantum gravity, which is not a  fully dynamical theory as in four dimensions but  is a theory where many computations can be done in detail. In particular, one should be able to  see in this theory exactly how noncommutative spacetime could emerge as a next-to-classical correction to conventional commutative spacetime. At the moment there are several candidate models for such noncommutative spacetimes even in the 3d setting.  Our goal in this paper is to bring all of these models into a single coherent picture, to explain precisely the relationships between the models at the structural level, and  to explore their physical implications to some extent. One important lesson we learn is that these relationships emerge only in the full theory {\em with} cosmological constant, as different degenerations related by a Hopf algebraic duality operation of `semidualisation'. Since we are mainly interested in the algebraic relationships we focus on the euclidean signature for simplicity, deferring the Lorentzian case to a sequel.

Of the various models, the most studied is the `spin model', which is  just the algebra of angular momentum but viewed as a noncommutative spacetime coordinate algebra. Its emergence as  an effective spacetime for 3d quantum gravity without cosmological constant was anticipated in \cite{thooft} and \cite{MatWell}. It was put forward 
 in \cite{BatMa:non} in view of its quantum symmetry group $D(U(su_2))$, whose role in 3d quantum gravity was proposed  in \cite{BM} and established in \cite{schroers}. The explicit emergence of this noncommutative spacetime starting from the Ponzano-Regge action was recently demonstrated in \cite{EL}. The $q$-deformation of this model, which, for $q$ a root of unity,  is the state sum behind the Turaev-Viro model,   describes 3d quantum gravity {\em with} cosmological constant as controlled by the quantum group $D(U_q(su_2))$. The $q$-deformed local spacetime here is the quantum group  $U_q(su_2)$ viewed as a noncommutative coordinate algebra. 

 Other models of  spacetime noncommutativity have been proposed,   which do not have  a firmly
 established  relation to quantum gravity.  In this paper we are particularly interested in 
  the  `bicrossproduct models'  introduced in the euclidean form in \cite{Ma:hop}  and in 3+1 form in \cite{MaRue:bic},  related to the construction of what was called $\kappa$-Poincar\'e symmetry  in \cite{LNRT:def}. The 3+1 bicrossproduct model is sometimes called `deformed special relativity'  but this is  misleading as there are several other deformations of special relativity under consideration, and  we therefore  keep the more specific name. This model is of particular interest because it predicts an energy dependent speed of light which will be tested by time of flight data currently being collected at the NASA Fermi gamma-ray space telescope (formerly GLAST) .
 Note, however, that there is little evidence of a theoretical link between the bicrossproduct model and quantum gravity.  In particular,  it was recently shown \cite{MSkappa}  that the 2+1 dimensional version of the bicrossproduct model (with a timelike noncommutative direction) does not arise directly in 3d quantum gravity. One of the upshots of the current paper is that  bicrossproduct models do have a precise role related to quantum gravity in its usual presentation, via our semidualisation map, or in physical terms by an interchange of position and momentum.  

Also in the 1990's there was completely developed a $q$-deformed Minkowski space theory in the form of $2\times 2$ braided Hermitian matrices \cite{Ma:exa}.  We will show that these various models are all intimately related. To do this we  use new results as well as  results known to experts in quantum groups, and explained, for example, in \cite{Ma:book}.  A subsidiary purpose of this paper is to advertise some of those results to the quantum gravity community, where they are not so well known (with notable exceptions, see e.g.  \cite{FNR}).

In order to  give an overview of our findings 
we need to look at the physical constants
that enter  quantum gravity, namely
the gravitational constant  $G$, Planck's constant $\hbar$  and the 
cosmological constant $\Lambda$ (we work in units where the speed of light is 1). In 3d gravity,
the dimension of $G$ is that of an inverse mass; the Planck mass is entirely classical 
and given by   
\bea
\label{planckmass}
m_p= \frac 1 G.
\eea
The cosmological constant has the dimension of inverse length squared, and can be
used to define a cosmological length scale $l_c$ via
\bea
\label{coslength}
l_c=\frac{1}{\sqrt{|\Lambda|}}.
\eea
A second length scale is given by  the Planck length, which takes the form
\bea
\label{lengths}
l_p=\hbar G= \frac{\hbar}{m_p}. 
\eea
The dimensionless parameter $q$ which plays the role of the deformation parameter
in this paper is related to the ratio of the two length scales $l_p$ and $l_c$.
More precisely it is given by 
\bea
\label{qfirst} 
 q=e^{-\hbar  G \sqrt{-\Lambda} }.
 \eea
 Note that this expression is specific to the euclidean theory we are considering in this paper;
 in the Lorentzian version one should replace $\Lambda $ by its negative in the above expression, 
 as explained in   \cite{sissatalk}.

In order to organise the various models and symmetries appearing in this paper, we begin
with the case where all three 
physical constants $\hbar, G$ and $\Lambda$ are non-zero.
The quantum group $D(U_q(su_2))$, with $q$ defined as in \eqref{qfirst},  plays an important role  in euclidean 3d quantum 
gravity with a non-vanishing cosmological constant  \cite{BNR}.
One can take  the limit $q\to 1$ in several  ways, with different physical interpretations.
 The first   is to take $\hbar \rightarrow 0$, keeping $G$ and $\Lambda$ fixed.  This gives an obviously  classical gravity  theory with cosmological  constant, so that  $\hbar=0$  
 but $l_c<\infty$ and $m_p<\infty$. 
 We will not be interested in this first limit, and will in fact set $\hbar=1$.
  A second way of taking the limit is  to let $G\rightarrow 0$, keeping $\hbar$ and $\Lambda$ fixed.
  This gives  a theory without gravitational self-interactions
 but with a cosmological constant,  so that  $l_c<\infty$ and $m_p=\infty$; the symmetry quantum group 
 of this model is $U(so_{1,3})$ and gravity is effectively a classical background on which a quantum particle propagates. 
A  third possibility  is to   take $\Lambda\rightarrow 0$ while keeping $\hbar$ and $G$ fixed, leading  to a  quantum gravity theory without cosmological constant i.e.   $m_p<\infty$ and $l_c=\infty$; the symmetry quantum group is now $D(U(su_2))$.
  The joint limit $G\rightarrow 0$ and $\Lambda \rightarrow 0$ with $\hbar \neq0$ 
   is a free quantum particle propagating  in euclidean space,  controlled by the group $E_3$ of euclidean motions.
   
   None of these limits  give the bicrossproduct models. Instead we need the semidualisation operation mentioned earlier. This comes out of quantum group theory and was used to understand both the quantum double and bicrossproducts. In general, semidualisation takes any quantum group built from factors (in our case momentum and rotations) acting on some other space (in our case position space) and swaps the roles of position and momentum. We will elaborate this in detail later, but for now we only need to know that an  original quantum group $H_1\dcross H_2$ acting on $H_2^*$ semidualises to a bicrossproduct quantum group $H_2^*\lrbicross H_1$ acting on $H_2$, assuming there is an appropriate notion of dual cf. \cite{Ma:phy,Ma:book}. It is important to note that not only do position and momentum get swapped, the quantum group also gets changed so this is a change of model and not merely a (quantum) Fourier transform of the same model.

\begin{table}
\begin{tabular}{|l|c|c|}
\hline
{\rm  3d  gravity \;\;\;\;\;\;\;\;} & $m_p=\infty$  & $m_p<\infty$\\
\hline
&&\\
$ l_c=\infty$ & $U(e_3)=U(su_2)\rcross \C[\R^3] $& $D(U(su_2))=U(su_2)\rcross\C[SU_2]$\\
$ l_c<\infty$ & $U(so_{1,3})=U(su_2)\dcross U(su_2^\star)$ & $D(U_q(su_2))\cong U_q(so_{1,3})$ \\
&&\\
\hline
\end{tabular}

\vspace{.5cm}

\begin{tabular}{|l|c|c|}
\hline
{\rm Semidual model} &$ m_p=\infty$ &$ m_p<\infty$ \\
\hline
&&\\
 $l_c=\infty$ &$ U(e_3)=  U(su_2)\rcross \C[\R^3]\;\;\;\;\;$ & $U(su_2)\tens U(su_2)=U(so_4)$\\
 $l_c<\infty$ & $ U(su_2)\rlbicross C[SU_2^\star]$ &$ U_q(su_2)\tens U_{q^{-1}}(su_2)=U_q(so_4)$\\
 &&\\
 \hline
 \end{tabular}
 
 \vspace{.5cm}

\caption{The quantum groups arising in 3d gravity for \hbox{$\Lambda \leq 0$},
Ä  and their semiduals. The diagonal entries are self-dual, up to a quantum Wick rotation in the $q\neq 1$ case.}
 \end{table}

The quantum groups arising  as limits  of $D(U_q(su_2))$ and their semiduals are listed in
Table~1 for $\Lambda\leq 0$, together with the physical regimes to which they are associated.
 The table also shows that the values of the physical constants associated to  semidual models are related by the exchange 
\bea
\label{dualswap}
m_p\leftrightarrow l_c.
\eea
Interestingly, this duality does not involve $\hbar$, since both $m_p$ and $\Lambda$
are purely classical.  Moreover, still assuming  $\Lambda\leq 0$, we note that 
we can write the deformation parameter  $q$ in \eqref{qfirst} as 
\bea
q=e^{-\frac{\hbar}{m_p l_c}}.
\eea
This is invariant under the duality \eqref{dualswap}. Thus, according to the table
quantum gravity with cosmological constant covered by $D(U_q(su_2))$ 
is in a certain algebraic sense  self-dual:  it is invariant under semidualisation  up to $q$-Wick rotation. 
This near self-duality is lost when one takes the limits $l_c\rightarrow \infty $ and $m_p\rightarrow
\infty$ separately, but reappears when  both limits are taken together:  
quantum theory of a free particle in  euclidean space
without cosmological constant, controlled  by $E_3$, is structurally  invariant under semidualisation and self-dual in this sense. Notice that  the requirement of self-duality or Born reciprocity requires  that  $m_p$ and $l_c$ are either both infinite    (the $E_3$ flat space model) or both finite  (the $q$-deformed model). Hence self-duality as an approach to quantum gravity, as advocated in \cite{Ma:pri},   forces the cosmological constant  to be non-zero.

Armed with this overview we can now outline the  paper. Section~2 contains background material on Poisson-Lie groups,  a  summary of the Chern-Simons formulation of 3d gravity and 
an  explanation of the concept of  semidualisation for  Hopf algebras.

Section~3  contains  a detailed explanation and elaboration of the structural relations  
between the Hopf algebras
summarised in  Table~1. We describe each of the Hopf algebras in detail, and give precise definitions
of the various limits, semidualisation maps and isomorphisms that relate them.  The general
statement, made earlier in this introduction,  that semidualisation swaps the role of positions and momenta is elaborated in this section, and illustrated by examples. 
An important role in this section is  played by  isomorphisms like the one   between $ D(U_q(su_2))$ and $U_q(so_{1,3})$
(indicated by $\cong$ in Table~1) which are  `purely quantum phenomena' in the sense that they only hold when $q\neq 1$.  
Taking the limit $q\rightarrow 1 $ on either side of such an isomorphism gives different quantum groups, and this provides 
the  mathematical  definition of   the physical distinction between taking the limit $l_c\rightarrow \infty$ and the limit
$m_p\rightarrow \infty$. 
A key finding of this section is  the result, already sketched above,  
that  3d quantum gravity with cosmological constant is self-dual
 up to $q$-Wick rotation. We also explain why this near self-duality   {\em fails }  in the limit  $q \rightarrow 1$. The  reason is 
that  a 'purely quantum' isomorphism  used in the near  self-duality
 breaks down when $q=1$ and therefore, when one takes the limit $q\rightarrow  1$,  
 one can do it on either side of the isomorphism,   and  will have different theories. 
 
In Section~4 of the paper we explore the physical meaning of semidualisation in greater detail. Ultimately we  would like to understand this operation and the `self-duality' under it in the full $q\neq 1$ 
 theory,  but the latter is at present too poorly understood at this level of detail. However, with the aid of noncommutative geometry we do obtain a clear picture in the degenerate cases.  Our starting point  for the physical interpretation in all cases is the fact that fundamental symmetries of physics enter quantum theoretical models via their representations. Thus the Klein-Gordon, Dirac and Maxwell equations all determine irreps of the Poincar\'e group, and the free Schr\"odinger equation determines an irrep of the Galilei group.   This applies in our models on the local `model spacetime' on which our quantum symmetry groups act, which is a part of the information in the theory  (it has to be supplemented by patching information according to the topology). Our strategy is therefore to study representations of  a model and its semidual, and to compare them.  

Next, we have said that semiduality interchanges position and momentum. So on the one hand we have particles moving on position space and forming a representation of our  quantum symmetry  group, and in the semidual model we have waves on what in the original model was called momentum space. We can use Fourier transform to map over the physics of the semidual model over to our original position space in order to compare with the original model, and we do this. Thus our original position space has two kinds of fields on it. One set are particles forming irreducible representations of the original quantum symmetry group and the other is a second set of fields forming an irreducible representation of the semidual  quantum symmetry group. Note that not only are position and momentum swapped under semidualisation but the quantum symmetry  group also changes as we have seen in Table~1.  Secondly, when the position space is classical but curved its Fourier dual is a noncommutative space, and vice versa, i.e. we need methods of quantum Fourier transform\cite{Ma:book} and noncommutative differential geometry in order to establish this picture.

It is instructive here to start with the trivial case of the group $E_3$, which we do in Section~4.1. The semidual theory is also controlled by $E_3$ but with position and momentum interchanged. The structure is self-dual in this sense, with duality implemented by the $\R^3$ Fourier transform, but of course the actual physics of interest is not.  Physical states are elements of irreps of $E_3$,
but are realised quite differently  on  the two sides of the semiduality. As expected, 
an irrep of $E_3$  on one side consists of waves in position space, obeying a first order differential constraint and  the wave equation.  But on the other side it consists of of monopole sections on spheres of increasing radius in position space. The two `physical models' here are equivalent under Fourier transform {\em and} an exchange of position and momentum. 
We express  the monopole sections in terms of a linear vector-valued function obeying
an algebraic constraint, and show that the algebraic constraint maps to the differential constraint under
Fourier transform. This itself is quite interesting and is explained in detail.

In Sections~4.2 and~4.3  we look at the similar semiduality  between  the $D(U(su_2))$ spin model (3d quantum gravity without cosmological constant) and a quantum particle on $SU_2$ with the action of $SU_2\times SU_2$ from the left and the right.  We start in~Section~4.2 with the $D(U(su_2))$ model and $U(su_2)$ as the noncommutative or `fuzzy' position space. The group $SU_2$ then plays the role of a curved momentum space. We show how to describe irreps of $D(U(su_2))$ in terms  of vector-valued functions on this (curved) momentum space, obeying an algebraic constraint. A quantum group Fourier transform \cite{Ma:ista,BatMa:non,FreMa:non} maps these to solutions of noncommutative wave equations. For spins $0,1/2$ and $1$ we recover the known \cite{BatMa:non} noncommutative wave equations on the spin-model noncommutative (`fuzzy') $\R^3$. Our approach can in principle be extended to obtain fuzzy wave equations of all spin.

Then, in Section~4.3,  we turn to the semidual model and write the irreps of  $SU_2\times SU_2$  in terms of  vector-valued functions on $SU_2$ (now interpreted as curved position space)  which 
obey a differential equation. This time, a noncommutative Fourier transform gives us a picture of the irreps for this model as noncommutative monopole sections on fuzzy spheres in noncommutative momentum space. The physics in this model is not the same as the physics in the previous model of which it is the semidual.  For example, the physical momentum values labelling the irreps are now discrete whereas before they were continuous. However, they have a `similar form' as a remnant of the near self-duality in the full $q$-deformed theory. 

This exemplifies the general construction. The semidual model, by construction, has its representations on a space which is the  (quantum Fourier or Hopf algebra) dual of the space where the original model has its representations (in the discussion above, the original model is represented on $H_2^*$ with Fourier dual $H_2$, which is the space where the semidual model is represented). So one always has one space where fields of both models live, which is functions on position space for one model and  functions on momentum space in the other.  In order to compare the two models further, we  fix the interpretation of this space, as fields on position space, say. Then irreps of  one (quantum) group are realised by means of a wave equation constraint and irreps of the semidual (quantum) group by means of an algebraic (projective module)  constraint. In the case of Sections~4.2 and~4.3 the space for one model is the angular momentum algebra and its dual  is that of functions on $SU_2$. However, unlike in the $E_3$ case, the (quantum) groups which are being represented in the two cases are different. Indeed, the models are different:
 one is quantum gravity 
 without cosmological constant and the other is a quantum particle with cosmological constant. 
In the $q$-deformed case we return to the quantum groups being algebraically (twisting) equivalent although still with different unitarity $*$-structure requirements. These remarks are developed further in our final Section~5.  The appendix contains a summary of 
facts about  forms and vector fields on  Lie groups in our conventions.

\smallskip 
 \noindent {\bf Remark on units.}  Most of this paper is concerned with quantum mechanical methods applied on classical backgrounds or in quantum gravity.  As a rule we therefore  set $\hbar =1$. To revert to physical units the reader should insert $\hbar$ every time a mass 
 is expressed in terms of an inverse length or a length in terms of an inverse mass. 
 
 \section{Background: 3d gravity and quantum groups}
 
 Here we provide the background in both physics and mathematics that we need for our analysis.
 After a short summary of   Poisson Lie group  theory we review  classical  3d gravity,  using the language of Poisson Lie groups. Motivated its role 3d quantum gravity,  
 we review the quantum double  and its properties. We  introduce the semidualisation functor  and study some of its properties.

\subsection{Poisson-Lie groups}
\label{PLreview}
We write $\g$ for the Lie algebra of a Lie group $G$. When we require explicit generators
we use a basis in  which the structure constants are purely imaginary.  In the case
of $G$ being unitary, this means that the generators are Hermitian, with real
eigenvalues, simplifying  our discussions of representation 
theory and quantum mechanics.  Additional results and conventions  regarding  the differential geometry of Lie groups, which are needed later in this paper, are  summarised   in Appendix~\ref{lieapp}. 
 
A Poisson Lie group means a Lie group $G$ which is a Poisson manifold, so there
is a Poisson bracket among smooth functions on $G$, such that the product map $G\times G\to G$ is a  map of Poisson spaces. Here $G\times G$ has the direct product Poisson-manifold structure. 
It is known that such a Poisson bracket is equivalent to a map $\delta: \g\to \g\tens \g$ at the Lie algebra level, called the Lie cobracket. It is just the adjoint of the Poisson bracket $\g^*\tens \g^*\to \g^*$ when restricted to $\g^*\subset C^\infty(G)$. The pair $(\g,\delta)$ with appropriate axioms is called a Lie bialgebra and should be thought of as an infinitesimal quantum group.  A Poisson-Lie group is quasitriangular if $\delta\xi=\ad_\xi(r)$ where $r\in \g\tens\g$ obeys the  classical Yang-Baxter equation and its symmetric part $r_+$ is ad-invariant. It is called {\em factorisable} if it is quasitriangular and $r_+$ is non-degenerate as a map g $\rightarrow $ g$^*$.  The associated Poisson-Lie group is similarly factorisable in this situation (either locally near the identity or, with appropriate technical assumptions, globally).
 For any Lie bialgebra there is a double $d(\g)=\g\dcross\g^{*op}$ which is factorisable as is its Poisson-Lie group $d(G)=G\dcross G^{*op}$ where $G^{*op}$ is the opposite (with reversed product) of the Lie group associated to the dual Lie bialgebra $\g^{*}$.  We will use $\star$ to denote the combination $*op$. This group and $G$ are both subgroups and the formula $su=(s\la u)(s\ra u)$ defines the `dressing action'  $\la$ of $G$ on $G^\star=G^{*op}$. The action $\ra$ the other way is called the `backreaction' or dual dressing action. These matters and the general $\dcross$ theory which they relate to were explained in \cite{Ma:mat}, where one of us proved a theorem that Lie splitting data  exponentiates whenever one factor is compact. This theorem holds for general factorisations not limited to the double or `Manin triple'.

Note that since $d(G)$ is factorisable, its dual $d(G)^*$ is a Poisson-Lie group that is  diffeomorphic 
 to $d(G)$, at least near the identity,   via a map 
 \[ Z:d(G)^*=G^*\codcross G\to d(G)\]
 given in this case canonically by multiplication in $d(G)$. Under this map orbits in $d(G)^*$ under the dressing action of $d(G)$ map over to conjugacy classes in $d(G)$ as spaces. We will use the symplectic structure on these orbits, which  are symplectic leaves for the Poisson bracket on $d(G)^*$. 

Quantum groups such as $\C_q[G]$ are quantisations of $G$ with its standard Drinfeld-Sklyanin Poisson bracket, defined for all semisimple Lie groups. Their duals $U_q(\g)$ deform the classical enveloping algebras $U(\g)$ and can also, with a bit of care, be viewed as quantisation of the Drinfeld dual $G^*$  \cite{STS}.
The quantisation of  $d(G)^*$  can be viewed as yielding $D(U_q(\g))$ i.e.  the quantum double construction for quantum groups to be described in detail later.

\subsection{Reminder of 3d gravity with point sources}

We  consider gravity in three dimensions coupled to matter in the form 
of a fixed number  of point particles, and review  the Chern-Simons formulation
of the theory.    For simplicity, we restrict attention to  three-dimensional  manifolds of the form $\Sigma\times \R$, where $\Sigma$ is a closed two-dimensional  manifold of 
genus $\gamma$ and with $n$ marked points, one for each point particle. Concentrating on the 
euclidean version,  we view gravity in a first order form  of a dreibein $e^a$, where $a=1,2,3$,  and a spin connection  $\omega$ with 
values in $so_3$.  These data can be combined together into a single $\gothg$-valued gauge field $A$, where $\gothg$ is  one of the following: the Lie algebra $e_3$
of the euclidean group $E_3$ (for vanishing cosmological constant), the Lie algebra $sl_2(\C)\cong so_{3,1}$
of $SL_2(\CC)$  
(for negative cosmological constant), and the Lie algebra $so_4$ of $SU_2\times SU_2$ (for positive cosmological constant). 
In the following we  write  $\CG$ for any
of the three associated simply connected  Lie  groups,  and  $\Lambda $ for the cosmological constant.  Introducing generators $P_a$ of translations
and generators $J_a$ of rotations, with commutation relations
\bea
\label{genrelns}
[J_a,J_b]=\imath\epsilon_{abc}J_c, \quad [P_a,J_b]=\imath\epsilon_{abc}P_c,\quad [P_a,P_b]=\imath\Lambda\epsilon_{abc}J_c,
\eea
the spin connection can be expanded $\omega=-\imath \omega_aJ^a$  and the gauge field $A$ is
\[
A=-\imath(e_aP_a+\omega_aJ_a).
\]
In order to define an action principle for this connection one requires a non-degenerate, invariant symmetric bilinear form  $k$
on the Lie algebra $\gothg$. In terms of the generators above this is given by 
\bea
\label{kform}
k(J_a,P_b) =- \frac{m_p}{8\pi}\delta_{ab},
\eea
with all other pairings of generators giving zero. The  standard Chern-Simons action for the connection $A$, formulated with the symmetric form  $k$,  then reproduces 
the Einstein-Hilbert action in the first order formalism, as  observed by 
Achucarro and Townsend \cite{AT} and elaborated  by Witten \cite{W}. 
The constant $m_p/(8\pi)$, which  is related to Newton's constant via  \eqref{planckmass},  is not normally
included in the symmetric form  $k$
 but instead kept as a coupling constant which multiplies the Chern-Simons
action.  However,   since the non-degenerate symmetric bilinear form ultimately determines
the Poisson structure on the phase space of the theory,   the inclusion of the physical constants here 
makes it easier to keep track of them in subsequent calculations.

The physical degrees of freedom of Chern-Simons theory  are  encoded in the $\CG$-valued holonomies of the connection $A$ as follows.
To each puncture $i$ we associate an element $\xi_i^*\in \gothg^*$ encoding the mass $\m_i$ and spin $s_i$ of the particle $i$  via
\[
\xi^*_i = \imath (\m_i P^*_3 + s_i J^*_3)
\]
in a dual basis.
Using the form \eqref{kform} we obtain an associated element  in $\gothg$:
\bea
\label{xipara}
\xi_i= -\imath \frac{8\pi}{m_p}(\m_i J_3+s_i P_3).
\eea
The curvature of the connection $A$ has a delta-function singularity at each  puncture $i$ with  coefficients lying in the 
adjoint orbit of the  $\xi_i$. Correspondingly, the holonomy around the puncture $i$ is forced to lie in the  
 conjugacy class  $C_i$ containing  $e^{\xi_i}$.
The extended phase space is
\bea
\label{extps}
 \tilde P=\CG^{2\gamma}\times\prod C_i
\eea
and the actual phase space is
\[ P=\{(A_\gamma,B_\gamma,\cdots A_1,B_1,M_i)\in \tilde P\ |\ [A_\gamma,B_\gamma^{-1}]\cdots [A_1,B_1]^{-1}\prod M_i=1\}/{\rm \Ad(\CG)}\]
The $A_i,B_i$ are holonomies around and through handles, while the $M_i$ are holonomies around our punctures, all with reference to some arbitrary base point $*$. 
The reader may wonder here where in the moduli space is the location of our $n$ marked points at any given time. The answer is that the physics is diffeomorphism invariant so to a large extent these are irrelevant. Correspondingly, all that we retain from $\Sigma$ in $P$ is its topology. However, one can say a bit more about ``positions'' of the particles in the theory. To do this  we need to consider the Poisson 
structure of the theory.

The gauge groups $\CG$ of the Chern-Simons formulation of gravity are all Poisson-Lie groups. 
The Poisson structure does not enter into the formulation of the gauge theory, but plays
an important role in describing the Poisson structure of its phase space, as we shall explain.
We focus on two here, both arising in the euclidean situation (later on we will suggest two more).
Without cosmological constant, we take 
\bea
\label{nocos}
\CG=d(SU_2)=SU_2\rcross su_2^*=E_3
\eea
 as a group but with a non-trivial Poisson bracket. Here $SU_2$ here is regarded as a Poisson-Lie group with the zero Poisson bracket, and we then take its double. Hence $su_2^*$ is a Lie algebra with zero Lie bracket and hence we can also view it as an abelian group, with the Kirillov-Kostant Poisson bracket. With  negative cosmological constant, we take 
\bea
\label{withcos}
\CG=d(SU_2)=SU_2\dcross SU_2^\star=SL_2(\C)
\eea
 as a group but with a non-trivial Poisson structure. Here $SU_2$ is a Poisson-Lie group equipped with its Drinfeld-Sklyanin bracket and we take its double.

There is a natural Poisson structure on $\tilde P$ given by a certain `braided tensor product' of those on each copy of $\CG\times \CG$ 
and on each conjugacy class \cite{FR}  which descends to the Atiyah-Bott one on $P$. In the Hamiltonian approach (see \cite{AGSI,AGSI,AS} and  \cite{BNR} in the context of 3d gravity), its quantisation  is the main step in constructing quantum gravity coupled to point sources. 
Equivalently the braidings can be untangled and $\tilde P$ 
is Poisson equivalent to the direct product of the Poisson structures on the conjugacy classes $C_i$ and the Heisenberg-double ones on 
$\gamma$ copies of $\CG\times \CG$ \cite{AM}. We  concentrate on the former, associated to the punctures.  The conjugacy classes $C_i$ in $\CG$ are the  image under a bijection 
\[
Z:\CG^*\rightarrow \CG,
\]
discussed in Section~\ref{PLreview},
 of the symplectic leaves of the Poisson-structure on $\CG^*$. The map is provided by  an invariant, non-degenerate symmetric
bilinear form  at the level of the associated Lie bialgebras (assuming again that we work with  the associated connected and simply connected Lie groups, or ignore certain global issues). 

To proceed further, we make use of the fact that the  Poisson-Lie groups discussed so far are all 
(special cases of) double cross products $\CG=G_1\dcross G_2$ of Poisson-Lie groups (this means that they factorise into the two Poisson-Lie subgroups and can be recovered from them by means of a double semidirect product in which each $G_1$ and $G_2$ acts on the set of the other and with the direct product Poisson structure).  Then $\CG^*=G_1^*\codcross G_2^*$ (a direct product as groups and a certain double-semidirect Poisson structure). One can describe the inverse images $Z^{-1}(C_i)$ in these terms. If  the Lie algebras
$\g_1$ and $\g_2$  of $G_1$ and $G_2$ have generators $J_a,P_a$ respectively (not necessarily the same as in \eqref{genrelns}), the dual Poisson-Lie group has Lie algebra  generators $J_a^*,P_a^*$, say, forming a dual basis to these  (so that $\< J^*_a, J_b\>
=\< P^*_a,P_b\>= \delta_{ab}$).
The coefficients in these bases form a local coordinate system for $\CG^*$ near the identity which we shall use, namely $j_a=\<-\imath J_a,(\ )\>$ is $-\imath J_a$ as linear functions on $\g_1^*$ and $p_a=\<-\imath P_a, (\ )\>$ as linear functions on $\g_2^*$. One may then write the Poisson bracket of $\CG^*$ explicitly among the $j_a$ and $p_a$. When restricted to $Z^{-1}(C_i)$ they form the classical phase space coordinates associated to each conjugacy class. 

Also, $\CG=G_1\dcross G_2$ acts canonically on the dual Poisson-Lie group $G_2^*$ (say) and one can form a cross product `Heisenberg-Weyl group'  $(G_1\dcross G_2)\rcross G_2^*$. In physics this group should be  represented in the quantum algebra of observables, i.e. its enveloping algebra as a quantisation of the dual Poisson-manifold $(G_1^*\codcross G_2^*)\rcocross G_2$ as an extended phase space. Here this copy of $G_2$ has coordinates near the identity which we denote now by $x_a=\<\imath P_a^*,(\ )\>$ as linear functions on $g_2$. One has then additional Poisson-brackets for these variables among themselves and with the previous $j_a,p_a$.  We shall prove these facts at the Hopf algebra level in Section~3 and the Poisson-Lie versions follow analogously.

To see all of this explicitly and also to understand the physical role of these `position variables' $x_a$,  we  concentrate  on the case of vanishing cosmological constant, so $\CG=E_3=SU_2
\rcross su_2^*$.  Our conventions for this group are spelled out in Section~\ref{e3sect}; note that they differ from
those used in a similar context in \cite{schroers} and \cite{MS1}.
The group $\CG^*$ is simply the direct product $E_3^*=su_2^* \times SU_2$ according to what we have said above. The map $Z$ is
\[ Z(\vec j,u)=(u,\Ad^*_u(\vec j)),\]
where we use our above bases for $\g_1^*$ and $\g_2$ 
in each case:  $j=\imath\vec j\cdot \vec J^*$ is an element of $g_1^*=su_2^*$ on the left and 
$-\tfrac{8\pi}{m_p}\imath \Ad^*_u(\vec j)\cdot \vec P$ is an element of  $g_2=su_2^*$ on the right\footnote{for the abelian Lie group $su_2^*$, the Lie algebra coordinates provide global coordinates on the group}. 
Meanwhile, the $P_a^*$ obey the rescaled $su_2$ commutation relations
\[  [P^*_a,P^*_b]=-\imath \frac{8\pi}{m_p}\epsilon_{abc}P^*_c.
\]
In view of the non-degenerate symmetric bilinear   form \eqref{kform} 
 on $e_3$ we could identify 
\bea
\label{pjrel}
 P^*_a \leftrightarrow
  -\frac{8\pi}{m_p} J_a,
\eea
 but will refrain from doing so to avoid confusion. 
Thus $u=e^{\imath p_aP_a^*}$ in terms of our local coordinates for $G_2^*$ near the identity. 

Let us focus on one conjugacy class $C$ containing the element $e^\xi$ with $\xi$
parametrised as in \eqref{xipara} (and the index $i$ dropped).
As we shall explain below,  one can describe the preimage  $Z^{-1}(C)$ of a conjugacy class $C$ in $\CG$
  as the subset of elements  $(j,u)\in \CG^*$ with coordinates obeying the further constraints
\bea 
\label{casimir} \vec p^2=\m^2,\quad \vec j\cdot \vec p=\m s .
\eea
 The Poisson  structure of $\CG^*$  gives rise to the brackets 
 \bea
\label{e3pbrackets}
\{j_a,j_b\}=\epsilon_{abc}j_c, \qquad \{j_a,p_b\}=\epsilon_{abc}p_c, \qquad \{p_a,p_b\}=0,
\eea
and it is easy to check that the combinations \eqref{casimir} are Casimirs, 
confirming that the conjugacy classes are indeed the symplectic leaves of the Poisson structure 
\eqref{e3pbrackets}.  
The Poisson brackets suggest that we should think of $p_a$ as the particle's momentum and 
$j_a$ as the particle's ``angular momentum'' coordinates. However, the coordinates $p_a$ fail
for the group element $u$ when $|\vec p|=m_p/4$ and $u=-1$. Thus, in 3d gravity we should really
interpret  $u$ as the particle's group-valued momentum. Momentum space is curved, and has the 
structure of a non-abelian Lie group. This is a classical effect, and means that, even
classically, momentum addition is noncommutative. 
 
 Geometrically, the space of vectors $\vec p$ and $\vec j$ obeying the constraints (\ref{casimir}) parametrise the  space of all lines in $\R^3$, and we shall see next that we may think of these
lines as the particle's world line in an auxiliary euclidean  space with the coordinates $x_a$. Thus, if we describe a symplectic leaf  of $\CG^*$ over in $\CG=E_3$ as a conjugacy class, we can redundantly parametrise it in terms of elements $(g,x)\in E_3$ that occur in 
$C=\{(g,x)^{-1}e^\xi(g,x)\}$.
The image under $Z$ of the point $(\vec j,u)$ in the physical phase space obeying (\ref{casimir}) maps over redundantly to a set of points $(g,x)\in E_3$ such that $Z(\vec j,u)=(g,x)^{-1}e^\xi(g,x)$. This set of points is described by  $g \in G_1=SU_2$ and  a coordinate vector $\vec x$  for  
$x=-\imath \vec x \cdot \vec P \in G_2=su_2^*$ obeying
\[ \vec j=\tfrac{m_p}{8\pi}(\Ad^*_{u^{-1}}-1)(\vec x)+ s \frac{\vec p}{\m},\quad \Ad_{g^{-1}}(\m  J_3)=\vec p\cdot \vec J.\]
Note that we have identified the translation part of the group $E_3$ with the position in
the auxiliary euclidean space by fixing an origin.
 The limit
\[
(\Ad^*(u^{-1})-1)\vec x\approx \tfrac{8\pi}{\m_p}\vec x \times \vec p
\]
for small $\m/m_p$ suggests, by analogy with the flat-space formula for angular momentum, 
 that we should interpret $\vec x$ as the particle's (spacetime) position. Further
support for this interpretation  comes from the following geometrical consideration. Position coordinates
should act on momentum space by translation. Since, as we just saw, momentum space is curved, such
translations cannot commute if they are to be globally defined.  One finds that 
\bea
\label{e3sklyanin}
\{x_a,x_b\}=-\tfrac{8\pi}{m_p}\epsilon_{abc}x_c,
\eea
as well as 
\bea
\{j_a,x_b\}=\eps_{abc}x_c,\quad \{x_a,f\}=-\tfrac{8\pi}{m_p}\xi^R_a(f)
\eea
for the Poisson brackets with the coordinates of $\CG^*$. Here $f$ is any function on $G_2^*=SU_2$ and $\xi^R_a$ is the right-translation vector field associated to the generator $J_a$ of the Lie algebra
according to \eqref{diffdef}. 
The geometrical meaning of these brackets is that the Poisson brackets of position coordinates are those 
of the $su_2$ Lie algebra, and that they act on the momentum manifold $SU_2$ as generators of 
right-multiplication. Note that the  bracket \eqref{e3sklyanin} is also part of the initial Poisson structure on $\CG= E_3$  (with all other brackets vanishing in our case). The conjugation action of $E_3$ on conjugacy classes is the dressing action on symplectic leaves of $\CG^*$; this is a Poisson action with the Poisson
structure of $\CG$ taken into account.

The above discussion  reveals  Poisson noncommutativity of  position coordinates
 in 3d gravity, 
but there are important caveats. First of all, we can change the coordinate vector
$\vec x$ to $\vec x+ \tau \frac{\vec p}{\m}$, 
where $\tau$ is an arbitrary real parameter, without changing the vectors
$\vec  p$ and $\vec j$. This is in agreement with our interpretation of $\vec p$ and $\vec j$ as parameters of 
a world line: shifting the position vector along the worldline does not change the worldline itself.
 The second, and more important, caveat is that  all of the above coordinates refer to the extended
phase space $\tilde P$ and are therefore not well-defined on the physical phase space $P$. One may
interpret them as referring to an auxiliary  euclidean space associated with the base point $*$ where the holonomies
start and end. However, to obtain the physical phase space we should divide by euclidean motions
in that  space. The Poisson brackets of physical quantities like traces of (products of) holonomies
have been studied in \cite{Martin}, but the relation with the above position coordinates has not 
been clearly established. An alternative approach is to study universes with boundary. In that case
there is a preferred family  of ``centre-of-mass frame'' of the universe. By choosing the 
base point to be associated with one such frame, the coordinates of the holonomies with respect
to the base point regain some of their physical meaning. This approach is pursued in \cite{MS1,MS4,MS5}.

The above description of the phase of 3d gravity in terms of the Poisson Lie structures associated to $\CG$ is tailor-made for the Hamiltonian approach to the quantisation of Chern-Simons theory \cite{AGSI,AGSII,AS}.  In this approach, a key role is played by a Hopf algebra $H$ which is a quantisation of the  Drinfeld dual $\CG^*$.  The Hilbert space of the quantised Chern-Simons theory can then be described in terms of representation theory of $H$ in a manner which is  analogous to the construction of the classical phase space as a quotient of the extended phase space \eqref{extps}.  Schematically (and referring to the above references for details) the quantisation of the extended phase space is a tensor product of $\gamma$ copies of a representation $\mathcal R$ of $H$,  which is the analogue of the regular representation of a group (and the quantisation of the Heisenberg double Poisson manifold $\CG\times \CG$),  and irreps $V_i$ of $H$ for each of the punctures (the quantisation of the conjugacy classes $C_i$). The Hilbert space of the quantised Chern-Simons theory is 
\bea
\label{Hilbertius}
{\mathcal H}=\text{Inv}\left({\mathcal R}^{\otimes \gamma}\otimes \;\;\bigotimes V_i\right),
\eea
where Inv denotes the $H$-invariant part of the tensor product. For the cases of euclidean gravity without  \eqref{nocos} and with negative cosmological constant \eqref{withcos},  the  relevant quantum groups  are the quantum doubles $D(U(su_2))$ and $D(U_q(su_2))$ ($q\in \RR$). Details of the Hamiltonian quantisation programme for these cases can be found, respectively, in \cite{MS2} and \cite{BNR}.

\subsection{Quantum double and semidualisation theorem}

Having motivated the role here of quantum groups in the picture, we now fix our notations for these, and recall the quantum double.  Let $H$ be a Hopf algebra over $\CC$, with coproduct $\Delta:H\to H\tens H$, counit $\eps:H\to \C$ and antipode $S:H\to H$. The particular real form of interest is expressed by, in addition, a $*:H\to H$ making $H$ into a Hopf $*$-algebra. We let $H'$ be a suitable dual of $H$ such that it is also a Hopf algebra and dually paired with $H$ by a non-degenerate map $\<h,a\>$. We refer to \cite{Ma:book} for all further details. It is useful to use the `Sweedler notation'  $\Delta h=h\o\tens h\t$.

The quantum double $D(H)=H\bowtie H'{}^{op}$ is built on the vector space $H\tens H'$ with new product
\[ (h\tens a).(g\tens b)=hg\t \tens b a\t \<g\o,a\o\>\<Sg\thr,a\thr\>,\]
 where $ h,g \in H, \; a,b \in H'$,  
and the tensor product coproduct \cite{Dri,Ma:phy}. This Hopf algebra has a canonical action \cite{Ma:book} on $H$
\[ h\la g= h\o g Sh\t,\quad a\la h=\<a,h\o\>h\t\]
and induces on it the canonical braid-statistics
\[ \Psi(h\tens g)=h\o g Sh\t\tens h\thr\]
with respect to which $H$ is $\Psi$-commutative. It also induces braid statistics on any other objects
covariant under $D(H)$. There is  a canonical action of $D(H)$ on $H$ which we can therefore view as a `noncommutative space'  (assuming the Hopf algebra $H$ is noncommutative). The dual of the quantum double is $H'\codcross H^{cop}$ which means  the tensor product as an algebra (its coproduct is twisted). It contains the `noncommutative position algebra' $H$ which ties in with our semiclassical picture above.

If $H$ is cocommutative i.e. $H'$ commutative we have $D(H)=H\rcross H'$ with
\[ (h\tens a)(g\tens b)=hg\o\tens a\ra g\t.b,\quad a\ra g=a\t\<g,a\o Sa\thr\>\]
i.e. the semidirect product by the right  coadjoint action corresponding left adjoint coaction of $H'$ on itself, see \cite{Ma:book} for the Hopf algebra formalism.

\subsection{Semidualisation}

The general construction of which the quantum double is part is a `double cross product' $H=H_1\bowtie H_2$ of a Hopf algebra factorising into two sub-Hopf algebras. Factorising means that the map $H_1\tens H_2\to H$ given by viewing in $H$ and multiplying there, is an isomorphism of linear spaces. In this situation one deduces actions $\la:H_2\tens H_1\to H_1$ and $\ra:H_2\tens H_1\to H_2$ of each Hopf algebra on the vector space of the other. These are defined  by $(1\tens a).(h\tens 1)=(a\o\la h\o\tens a\t\ra h\t)$ for the product of $H$ viewed on $H_1\tens H_2$. The coproduct of $H_1\dcross H_2$ is the tensor one given by the coproduct on each factor and there is a canonical action of this Hopf algebra on  the vector space of $H_1$ by
\[ (h\tens a)\la f=h. (a\la f),\quad \forall f\in H_1,\quad h\tens a\in H_1\tens H_2.\]
This in fact respects the coalgebra structure of $H_1$ and hence provides in a canonical way a covariant right action of $H_1\dcross H_2$ on $H_1'$ as an algebra. Explicitly, the right action of $H_2$ on $H_1'$ is defined by
\[ \<\phi\ra a,h\>=\<\phi,a\la h\>,\quad \forall \phi\in H_1',\ a\in H_2,\ h\in H_1, \]
and in these terms the right action of $H_1\dcross H_2$ on $H_1'$ is 
\[ \phi\ra(h\tens a)=\<\phi\o,h\>\phi\t\ra a.\]
In this case we may form the cross product algebra by this action
\bea
\label{A1}
(H_1\bowtie H_2)\rcross H_1'.
\eea

Also in this situation we may dualise one of the factors, say replacing $H_1$ by  $H_1'$. This gives a new Hopf algebra $H_2\rlbicross H_1'$ (the semidual of $H$) which then acts covariantly from the left on $H_1$ as an algebra. The product and coproduct are
\[ (a\tens\phi)(b\tens\psi)=ab\o\tens \phi\ra b\t\psi,\quad \Delta (a\tens \phi)=(a\o \tens a\t\bz \phi\o)\tens(a\t\bo\tens \phi\t)\]
\[a\ra h=\<a\bz,h\>a\bo,\quad h\in H_1,\ a,b\in H_2,\ \phi,\psi\in H_1', \]
where the coaction on $a\in H_2$ is defined in terms of our original $\ra$. Its canonical left action on $H_1$ is
\[ (a\tens\phi)\la h=a\la h\o \<\phi,h\t\>.\]
This is the `semidualisation functor' that applies to Hopf algebras that factorise \cite{Ma:phy,Ma:book}. In this case we have a cross product algebra by the action on $H_1$:
\bea
\label{A2}
H_1\lcross (H_2\rlbicross H_1').
\eea

\begin{lemma}
\label{semilemma}
 The two algebras \eqref{A1} and \eqref{A2} are the same when built in the vector space $H_1\tens H_2\tens H_1'$ Hence there is one algebra
\[
A=(H_1\bowtie H_2)\rcross H_1'=H_1\lcross (H_2\rlbicross H_1')
\]
 independently of the point of view, with
\[ H_1\bowtie H_2\subset A\supset H_2\rlbicross H_1'\]
i.e., containing both the double cross product and the bicrossproduct. Moreover, $A\supset H_1\rcross H_1'=H_1\lcross H_1'$ the Heisenberg-Weyl algebra.
\end{lemma}
\proof This is automatic from the definition of the semidualisation process when one goes into how this is actually defined by dualising the involved actions and coactions. Indeed, the product of $A$ computed the first way is
\[ (h\tens a\tens\phi).(g\tens b\tens \psi)=(h\tens a).(g\tens b)\o\tens( \phi\ra (g\tens b)\t).\psi\]
\[ \quad=(h\tens a).(g\o\tens b\o)\tens\<\phi\o,g\t\>(\phi\t\ra b\t).\psi\]
\[ \quad =h.(a\o\la g\o)\tens (a\t\ra g\t).b\o\tens \<\phi\o,g\thr\>(\phi\t\ra b\t).\psi.\]
Meanwhile, computing the product the other way gives
\[ (h\tens a\tens\phi).(g\tens b\tens \psi)=h.((a\tens \phi)\o\la g)\tens (a\tens\phi)\t.(b\tens\psi)\]
\[ =h.( (a\o\tens a\t\bz\phi\o)\la g)\tens(a\t\bo\tens\phi\t).(b\tens\psi)\]
\[ =h.(a\o\la g\o)\<a\t\bz\phi\o,g\t\>\tens a\t\bo b\o\tens (\phi\t\ra b\t).\psi\]
\[ =h.(a\o\la g\o)\<a\t\bz,g\t\>\<\phi\o,g\thr\>\tens a\t\bo b\o\tens (\phi\t\ra b\t).\psi,\]
which is the same on using the definition of the coaction on $H_2$. Also, the product restricted to $h\tens\phi=h\tens 1\tens\phi$ is 
\[ (h\tens\phi).(g\tens\psi)=hg\o\tens\<\phi\o,g\t\>\phi\t\psi,\]
which can be viewed either way   $H_1\rcross H_1'=H_1\lcross H_1'$ as a cross product of the coregular representation (in the finite dimensional case it is the matrix algebra $End(H_1)$\cite{Ma:book}. \endproof

This gives a concrete rotation-momentum-position algebra way of thinking about semidualisation. The three form a single algebra.  If we think of $H_1,H_2$ as momentum, rotations we see the double crossproduct acting on $H_1'$ as positions, and if we think of  $H_2,H_1'$ as rotations, momentum, we see the bicrossproduct acting on $H_1$ as positions. This is a version of `quantum Born reciprocity' (interchanging position and momentum) which is a little different from the original motivation for bicrossproducts as quantum phase spaces, but based on entirely the same Hopf algebra dualisation constructions namely to interchange $H_1$ with $H_1'$. If one looks only at the position-momentum sector then this is the usual Heisenberg-Weyl algebra (sometimes called the `Heisenberg double') which is symmetric between position and momentum so already admits the quantum Born reciprocity.

Finally, we can do the exact same constructions with the roles of $H_1,H_2$ swapped. Thus, there is similarly a canonical right action of $H_1\dcross H_2$ on the coalgebra of $H_2$ and its dualisation is a canonical left action on the algebra of $H_2'$. We can form a cross product by this. Alternatively, we can use the left action of $H_1$ on $H_2'$ and a right coaction of $H_2'$ on $H_1$ corresponding to $\la$ to define a bicrosspropduct $H_2'\lrbicross H_1$ which acts from the right on the algebra of $H_2$. As before, we have
\[ B= H_2'\lcross (H_1\dcross H_2)= (H_2'\lrbicross H_1)\rcross H_2\]
as an algebra
\[ H_1\dcross H_2\subset B\supset H_2'\lrbicross H_1\]
within which the semidualisation takes place. It contains $H_2'\lcross H_2=H_2'\rcross H_2$. We will actually use the $A$-version of semidualisation, given in  Lemma~\ref{semilemma},   in order that the bicrossproducts act naturally from the left, but this means that the double cross product acts naturally from the right. In the primary 3d quantum gravity models we prefer the $B$-version so that the double acts naturally from the left, but then the bicrossproduct acts from the right. To study their semiduals we flip conventions and use the $A$-model so that it is the bicrossproducts which act from the left (this is because physicists
tend to avoid right actions in actual computations).

In particular, if one applies the second version of the semidualisation (with dualising algebra $B$)  to $D(H)=H\bowtie H'{}^{op}$ one has the canonical Schr\"odinger left action on $(H'{}^{op})'=H^{cop}=H$ as an algebra as mentioned above. According to the above, we also have
\[ B=H\lcross D(H)=M(H)\rcross H'{}^{op}\isom (H^{cop}\tens H)\rcross H'{}^{op}\]
for some `mirror product' bicrossproduct
\[ M(H)=H^{cop}\lrbicross H\isom H^{cop}\tens H, \]
which  as stated turns out to be isomorphic to the tensor product Hopf algebra \cite{Ma:phy}. In effect, the quantum Mach principle or semidualisation (used the other way) converts something trivial over to something non-trivial, namely the quantum double, and was our way to construct it. The action of $H^{cop}\tens H$ on $H'{}^{op}$ from the right is
\[ a\ra (h\tens g)=\<h,a\o\>a\t \<Sg,a\thr\>\]
when one traces through the explicit constructions and isomorphisms. Note that $H\subseteq D(H)$ appears in $H^{cop}\tens H$ embedded on the diagonal via the coproduct. Its right action is therefore evaluation against the left adjoint coaction of $H'$ on itself. Likewise, if we use the $A$-version in order to have a left action here, and start with $D(H)=H'{}^{op}\dcross H$ acting from the right on $H^{cop}$ as an algebra, then the semidual is $H\rlbicross H^{cop}\isom H\tens H^{cop}$ acting on $H'{}^{op}$ from the left by
\[  (h\tens g)\la a=\<Sh,a\o\>a\t \<g,a\thr\>.\]
All operations in these formulae refer to the underlying Hopf algebra $H$ or its dual.

\section{Structure of the models as limits of 3d quantum gravity}

After  the above background, we describe in detail potentially eight  noncommutative spacetime models for the eight entries in Table~1. At this stage we are interested in the structure  of 
the symmetry algebras  of the models  and at this level describe   isomorphisms which reduce our models to only six. The more detailed situation is shown in Figure~1, as we shall explain in this section.

\begin{figure}
 \[
\includegraphics[scale=.9]{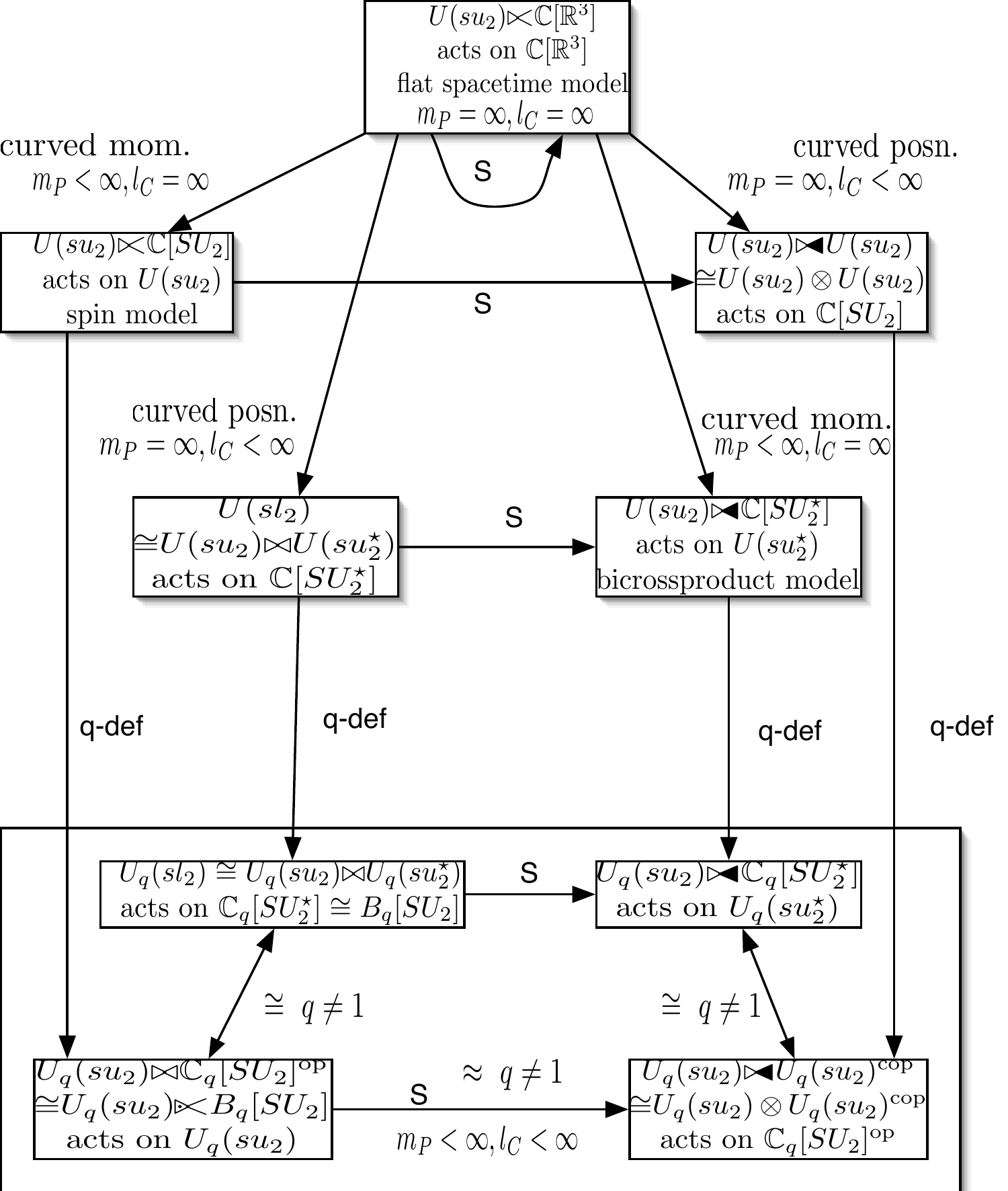}\]
\caption{Overview of isometry quantum groups in euclidean 3d quantum gravity models (left) and their semiduals (right). Here $SU_2$ is a 3-sphere,  $SU_2^\star$ is hyperbolic space, $U(su_2)$ and $U(su_2^\star)$ are noncommutative versions of $\R^3$. We denote semidualisation by $S$. }
\end{figure}

We will also introduce explicit notations for our examples. We clarify first an important piece of notation. 
In physics, the word momentum can be used in two ways: (a) with reference to a point in momentum space $\vec p\in\tilde\R^3$, or (b)  as an observable, which means its components $P_a$ are particular functions {\em on} momentum space. When Lie symmetries are realised they usually appear in the second form. For example $U(\R^3)$, with generators $P_a$ acting  on the algebra $\C[\R^3]$ of functions on position space by $P_a=-\imath{\partial \over \partial  x_a}$, is also  the polynomial algebra $\C[P_1,P_2,P_3]=\C[\tilde\R^3]$ of functions on momentum space. In this point of view,  $P_1$ is an infinitesimal element of $\R^3$ in the direction $(1,0,0)$ etc; it is a tangent vector in the Lie algebra of $\R^3$ and not a function on it.  Rather, each $P_a$ is a function on $\tilde\R^3$. This is clearer perhaps in the non-abelian case where $U(\text{g})$ acts naturally by vector fields on $C^\infty(G)$, so elements of g here are tangent not cotangent vectors. At the same time, they are functions on cotangent space. Finally, although we will not often make this distinction, one can think of $\vec p=(p_1,p_2,p_3)$ not as an actual numerical point but as a generic point, i.e. as a place-holder for actual but unspecified points  in momentum space $\tilde\R^3$. As soon as one does this, $p_a$ becomes a coordinate function on $\tilde\R^3$, i.e. acquires the same status as $P_a$. Thus, it will often be useful (and would be normal in physics) to mix notations in this way in order to avoid a proliferation of symbols.
  
\subsection{$E_3=SU_2\rcross\R^3$ -- free particle without cosmological constant  (flat spacetime)}
\label{e3sect}
We actually work with the double cover of the euclidean group of 
motions in three dimensions:
\bea
\label{E3group}
E_3=SU_2\ltimes \RR^3,
\eea
where we view $SU_2$ with the {\em zero Poisson bracket} and  $\R^3$ denotes the translation group with zero Lie bracket {\em and} zero Poisson bracket. The vanishing of the Lie bracket (commutativity of spacetime translations) amounts to taking the 
cosmological constant to be zero (or, by \eqref{coslength},  $l_c=\infty$) and 
the zero Poisson bracket on $E_3$ corresponds to a vanishing gravitational coupling constant (or, by \eqref{planckmass},
 $m_p=\infty$).
 The action of $SU_2$
is by rotations which can be expressed concisely as
\bea
\label{e3multi1}
(g,a)(h,b)=(gh, \Ad^*_h(a)+b),\quad g,h\in SU_2,\quad a,b\in su_2^*, 
\eea
where we identify our abelian translation group as $\R^3=su_2^*$. We denote as before the generators of $su_2^*$ by $P_a$.
We assume these generators to be proportional to the duals $J^*_a$ of the $su_2$ generators $J_a$, but not necessarily equal
to them. The reason for this is that different normalisations of the $J_a$ relative to the $P_a$ are required in different 
contexts, see e.g. \eqref{kform} in relation to 3d gravity. The upshot is that the $P_a$ form an orthogonal basis of $su_2^*$
 and that an element $a\in su_2^*$ can be written in terms of 
 a coordinate vector $\vec a$ as $a=-\imath \vec a\cdot \vec P$ in our conventions. 
The coadjoint action here  is a right action  defined by $\Ad^*_h(a): k \mapsto a(h(k )h^{-1})$, for $k\in su_2$,  
which we can also write by abuse of notation as $\Ad_{h^{-1}}(\vec a)$.  
In terms of the coordinate vectors $\vec a,\vec b\in\R^3$
for the $su_2^*$-elements $a$ and $b$ the above multiplication law is thus
\bea
\label{e3multi2}
(g,\vec a)(h,\vec b)=(gh, \Ad_{h^{-1}}(\vec a)+\vec b). 
\eea
By definition we also view the generators $P_a$ as coordinates on momentum space, generating its commutative coordinate algebra. The momentum space itself is the Lie algebra $su_2$ as another copy of $\R^3$. 

The Lie algebra $e_3=su_2\rcross \RR^3=su_2\rcross su_2^*$ has rotation generators $J_a$ and translation generators $P_a$
with commutation relations
\bea
\label{e3relns}
[J_a,J_b]=\imath\epsilon_{abc}J_c, \qquad [P_a,J_b]=\imath\epsilon_{abc}P_c,\qquad [P_a,P_b]=0.
\eea
Note that this Lie algebra   is {\em not} a classical double since $su_2^*$ here has the zero Lie cobracket, and its enveloping algebra  $U(e_3)=U(su_2)\rcross U(su_2^*)=U(su_2)\rcross\C[\tilde\R^3]$ is {\em not} a quantum double. It is, however, still an example of our more general double cross product. Hence there is a canonical action on the position space algebra  $\C[\R^3]$. It is  the local spacetime in the model and we see that it is flat.  Explicitly, the actions of the Lie algebra generators on 
 scalar functions $f(\vec x)$ on position space are defined by
\bea
\label{e3gens}
 P_a=-\imath {\del\over\del x_a},\quad J_a=-\imath\eps_{abc}x_b{\del\over\del x_c}.
 \eea
The physics which this theory describes has a model spacetime flat $\R^3$, which means that in each patch of $\Sigma\times\R$ the motion is that of a free particle on $\R^3$. There can still be a nontrivial $e_3$ connection which  is now everywhere flat regardless of the matter, but has nontrivial transitions between patches. Thus the particles respond to the background geometry but they do not act as sources for it. In short the model is the quantum theory of a particle on a flat background, possibly nontrivial.

The semidual model with flipped conventions is  given by
\[  E_3=SU_2\ltimes \tilde\RR^3
\]
where $\tilde\R^3$ has  zero Lie bracket and zero Poisson bracket, which we identify with $su_2$ as a vector space.  Its enveloping algebra is $U(su_2)\rcross U(su_2)=U(su_2)\rcross \C[\R^3]$  and acts naturally on the momentum coordinates $\C[\tilde\R^3]$. Clearly we can Fourier transform from functions on $\R^3$ to functions on $\tilde\R^3$ and back and thereby convert a construction in one model to one on the other where it will have a different interpretation. The algebraic structure, however, is self-dual under semidualisation. 

\subsection{$D(U(su_2))$ -- quantum gravity without cosmological constant (spin spacetime)}
\label{spinspace}
Next we take $SU_2$ with its {\em zero Lie cobracket} and $su_2^*$ the dual Lie bialgebra, which means
with the zero Lie bracket and Kirillov-Kostant Lie cobracket.  The classical Poisson Lie group is the double $d(SU_2)=SU_2\rcross su_2^*=E_3$ again but this time with a non-trivial Poisson-bracket. Its quantisation is the quantum coordinate algebra of the quantum symmetry group $D(U(su_2))=U(su_2)\rcross \C[SU_2]$, where $\C[SU_2]$ is the coordinate algebra on the momentum space $SU_2$ and is described by a matrix of generators $t^i{}_j$ dually paired with generators $J_a$ of  $U(su_2)$ by 
$\<t^i{}_j,J_a\>={1\over 2}\sigma_a{}^i{}_j$. Here $\sigma_a$ are the usual  Pauli matrices 
\bea
\label{pauli}
\sigma_1 =\bpm 0 & 1 \\ 1 & 0 \epm, \qquad 
\sigma_2 =\bpm 0 & -\imath \\ \imath & 0 \epm, \qquad 
\sigma_3 =\bpm 1 & 0 \\ 0 & -1 \epm. 
\eea

We describe the quantum double here in an algebraic form and with a parameter $\lambda$ that expresses the `flattening' of the momentum space $SU_2$ to $\R^3$ as $\lambda\to 0$. In the context of  3d quantum gravity one should take $\lambda=1/m_p$.  The algebraic quantum double then has generators $J_a$ of $su_2$ and generators  $t^i{}_j$ of the coordinate algebra  of  $SU_2$ with relations
\[ [t^i{}_j,J_a]={1\over 2}(\sigma_a{}^i{}_l t^l{}_j-t^i{}_l\sigma_a{}^l{}_j),\quad [J_a,J_b]=\imath\eps_{abc}J_c\]
\[ \Delta J_a=J_a\tens 1+1\tens J_a,\quad \Delta t^i{}_j=t^i{}_l\tens t^l{}_j.\]
We now change variables from $t^i{}_j$  to $\P_0,\P_1,\P_2,\P_3$ defined   via  
\[  t^i{}_j=\P_0\delta^i{}_j+\imath{\lambda\over 2} \P_c\sigma_c{}^i{}_j=\begin{pmatrix}\P_0+\imath {\lambda\over 2} \P_3& \imath{\lambda\over 2}( \P_1-\imath  \P_2)\\  \imath{\lambda\over 2} (\P_1+\imath \P_2)& \P_0-\imath {\lambda\over 2} \P_3\end{pmatrix}.\]
The structure in terms of the new generators is
\[ \P_0^2+{\lambda^2\over 4}\vec\P^2=1,\quad [\P_0,J_a]=0,\quad [\P_a,J_b]=\imath\eps_{abc}\P_c,\]
\[  \Delta \P_0=\P_0\tens \P_0-{\lambda^2\over 4}\P_a\tens \P_a,\quad \Delta \P_a=\P_a\tens 1+1\tens \P_a-{\lambda\over 2}\eps_{abc}\P_b\tens \P_c, \]
where the $\det t=1$ relation appears now as the sphere relation for $SU_2$ as a three-sphere in $\R^4$, with the $\vec\P$ the local coordinates of a patch of $SU_2$ containing the group identity. Here $\P_a$ are regarded as the free variables valid for $|\vec\P|\le 2/\lambda$ and $\P_0=\sqrt{1-{\lambda^2\over 4}\vec\P^2}$ in this patch. There is another patch covering the lower half with $\P_0\le 0$. In either patch, we see that $SU_2$ as momentum space for this model is a curved version of $\R^3$ obtained in the limit $\lambda\to 0$. Note that the two patches above are not open sets, one should really use open patches and a third patch around the equator to see the limit toplogically. 

We have a canonical action of the quantum double on $U(su_2)$  which means on the flat but noncommutative spacetime algebra
\bea
\label{ncspace}
 [x_a,x_b]=\imath \lambda \eps_{abc} x_c,
\eea 
where we recall that $\lambda$ is $1/m_p$ i.e. proportional to 
the Planck length $ l_p$ in the context of 3d gravity \eqref{lengths}. 
This is the enveloping algebra $U(su_2)$ with rescaled generators. The action of the quantum double  on the 
$x_a$ is
\[ J_a\la x_b=\imath\eps_{abc}x_c,\quad \P_0\la x_a=x_a,\quad \P_a\la x_b=\imath\delta_{ab},\]
see \cite{BatMa:non}.  

Finally, the $\vec \P$ coordinate system on momentum space $SU_2$ can be replaced by a local coordinate system $\vec p$ valid near the group identity. Here an element of $SU_2$ is written as $e^{{\imath\over 2}\lambda \vec p\cdot \vec \sigma}$ in terms of a vector of Pauli matrices and valid for $|p|< 2\pi/\lambda$. The relation between the two coordinate systems is 
\bea
\P_a=p_a{\sin({\lambda|\vec p|/ 2})\over{\lambda |\vec p|/ 2}},\quad \P_0=\cos({\lambda|\vec p|/ 2}). \nonumber
\eea
Note that this second `Lie algebra' coordinate system is degenerate at $|\vec p|=2\pi/ \lambda $ as all directions of $\vec p$ then lead to the same point $-1\in SU_2$.  The noncommutative geometry of the model can be considerably developed \cite{Ma:spo}\cite{FreMa:non}. In particular, in any reasonable completion of the position coordinate algebra to include exponentials, the elements $\zeta=e^{\imath\vec p\cdot \vec x}$ with $|\vec p|=2\pi/\lambda$ are non-trivial plane waves (of momentum $-1$) obeying $\zeta^2=1$ \cite{FreMa:non}.  This means that noncommutative spacetime is a kind of double cover of noncommutative $\R^3$ in the same way that $SU_2$ is a double cover of $SO_3$.  

This model describes quantum gravity without cosmological constant  in the sense that compared to the model of Section~3.1 the particles at each puncture of $\Sigma$ act as sources for the implicitly defined `connection'. This is achieved by switching on a finite $m_p$ or nonzero Newton constant $G$. The model spacetime is noncommutative and the  `connection'  is implicitly defined by its quantum group `holonomy' so is in that sense  `quantum'.   It is actually the combination $l_p=\hbar/m_p$ that enters so one could view the model equivalently as switching on $\hbar$ for fixed  $G$. In this way the theory describes quantum gravity coupled to the sources in contrast to Section~3.1 where the background geometry on $\Sigma\times\R$ remains classical and unaffected by the sources.

\subsection{$\widetilde{SO_4}$  -- free particle with positive cosmological constant ($SU_2$ spacetime) as semidual of quantum gravity without cosmological constant}

Next we apply the semidualisation construction to the previous quantum double spin model. Due to our analysis for any quantum double we obtain, in the present case, the quantum group  
\[U(su_2)\rcross U(su_2)^{cop}\isom U(su_2)\tens U(su_2)^{cop}=U(su_2\oplus su_2)=U(so_4),
\]
which is actually a classical enveloping algebra, acting covariantly on the classical position algebra  $\C[SU_2]$ by left and right translation. Note that in terms of the generators of rotations and `translations' on the left we have commutation relations
\bea
\label{jpcoms}
 [J_a,J_b]=\imath\epsilon_{abc} J_c,\quad  [J_a,P_b]=\imath\epsilon_{abc} P_c,\quad  [P_a,P_b]=\imath\lambda\epsilon_{abc} P_c,
\eea
where in this model $\lambda=1/l_c$. Its action on $\C[SU_2]$ is with $J_a$ acting as the vector fields for conjugation and $P_a$ acting as the vector fields for right translation. We can choose coordinates on $SU_2$ with parameter $\lambda$ as in Section~3.2, just now the $SU_2$ is position space, with $P_a$ becoming usual differentiation on flat $\R^3$ as $\lambda\to 0$. This represents a fairly perverse but physical way of thinking about left and right translations on $SU_2$  which we will develop further.

We see that the semidual of our flat but noncommutative spacetime and quantum gravity system is a system with curved but classical model spacetime $SU_2$.  At the group level the euclidean group is now deformed to $SU_2\rcross SU_2$ which is isomorphic to $SU_2\times SU_2$ and we view this as a double cover $\widetilde{SO_4}$.  In terms of the  notation  \eqref{diffdef},
the left copy of $SU_2$ acts by the vector fields $\xi^L$ and the right copy by the vector fields $\xi^R$ on functions on the position space $SU_2$. The theory deforms the flat model of Section~3.1 in now describing a quantum particle on a classical background with curvature (due to the cosmological constant) but insensitive to the sources. The motion looks locally like free motion on 3-spheres in each patch of $\Sigma\times\R$ with $\widetilde{SO_4}$ transitions.

This model is not self-dual as it is clearly very far from the previous model in Section~3.2. Thus, a  construction in quantum gravity but without cosmological constant maps over under semidualisation to a construction on classical $SU_2$. In physical terms of the original model this $SU_2$ is the curved momentum space. In the dual theory it is the curved position space. Conversely, a classical particle in the semidual theory means  a particle on $SU_2$ with $SU_2\times SU_2$ isometry group. It maps back to something else in the noncommutative spacetime of the quantum gravity model. We shall give details of both sides in Section~4.

\subsection{$\widetilde{SO_{1,3}}$ -- free particle  with negative  cosmological constant (hyperbolic spacetime)}

Here we take, in place of $E_3$, the classical group
 \[ SL_2(\C)=SU_2\dcross SU_2^\star\]
but with the zero Poisson bracket. Its structure is a double cross product of $SU_2$ and a certain solvable group $SU_2^\star=\R^2\lcross\R$ occurring in the Iwasawa factorisation. Each element of $SL_2(\C)$ may be uniquely factorised in the form
\[ \begin{pmatrix}a&b\\ c&d \end{pmatrix}=\begin{pmatrix}x & -\bar y \\ y & \bar x\end{pmatrix}\begin{pmatrix}w &z \\ 0 & w^{-1}\end{pmatrix},\quad |x|^2+|y|^2=1,\quad w>0,\quad x,y,z\in\C.\]
Such a matrix is in $SL_2(\C)$ and, conversely,  given a matrix as on the left, we define
\[ w=\sqrt{|a|^2+|c|^2},\quad x=w^{-1}a,\quad y=w^{-1}c,\quad z=w^{-1}(\bar a b+\bar c d).\]
Note that the group $SU_2^\star$ and the Iwasawa factorisation can be understood in Poisson-Lie terms  \cite{Ma:mat}. Thus, the former is the dual of $SU_2$ as a Poisson-Lie group with its Drinfeld-Sklyanin Poisson bracket and $SL_2(\C)$ is  the classical double of $SU_2$ as a Poisson-Lie group, but in the present model we use only the resulting $SL_2(\C)$ group and factorisation structure, taking it  with zero Poisson structure. 

There is a canonical right action of $SL_2(\C)$  from the classical group double cross product theory on the set $SU_2^\star$ as a classical but curved position space, 
\[ b\ra (g\dcross a)=(b\ra g).a.\] Using the above we can compute $\ra$ explicitly as
\[ \begin{pmatrix}w &z \\ 0 & w^{-1}\end{pmatrix}\ra \begin{pmatrix}x & -\bar y \\ y & \bar x\end{pmatrix}= \begin{pmatrix}w' &z' \\ 0 & w'{}^{-1}\end{pmatrix}\]
\[ w'=\sqrt{w^{-2}|y|^2+|wx+zy|^2},\quad w'z'=(w\bar x+\bar z\bar y)(z\bar x-w\bar y)+w^{-2}\bar x\bar y. \]
In this way  $SL_2(\C)$ becomes the isometry group of this position space with its natural hyperbolic metric, and the double cross product structure exhibits it explicitly as a curved position space analogue of the euclidean group of motions. $SU_2$ acts as `deformed rotations' $\ra$ and `deformed momentum' $SU_2^\star$ acts by group right-translation. In its internal structure $SU_2$ also acts on momentum by the same deformed action $\ra$ but as $SL_2(\C)$ is not a semidirect product, there is also a back-reaction 
\[ \begin{pmatrix}w &z \\ 0 & w^{-1}\end{pmatrix}\la \begin{pmatrix}x & -\bar y \\ y & \bar x\end{pmatrix}=w'^{-1} \begin{pmatrix}wx+zy &-w^{-1}\bar y\\ w^{-1}y & w\bar x+\bar z\bar y\end{pmatrix}\]
of momentum on rotations as a result of the curved space.   

At the algebraic level we have a left action of $U(sl_2)=U(d(su_2))=U(su_2)\dcross U(su_2^\star)$ on $\C[SU_2^\star]$ as the commutative coordinate algebra of functions on the classical but curved position space $SU_2^\star$. Explicitly, the generators of $sl_2$ as isometry Lie algebra are $J_a$ as usual for rotations and $P_a$, say, for `translations', with nonzero commutation relations
\begin{align}
\label{jphyp}
[J_a,J_b]&=\imath\eps_{abc}J_c, \quad   [P_3,P_i]=\imath\lambda P_i,\\
 [J_a,P_b]&=\imath\eps_{abc}P_c+\imath\lambda\delta_{b3}J_a-\imath\lambda\delta_{ab}J_3, \nonumber 
\end{align}
where $i=1,2$ and $\lambda=1/ l_c$ in this model. The parameter ensures that we recover $e_3$ as $\lambda\to 0$. Note that the quantum group in this example is a classical enveloping algebra and therefore is {\em not} a quantum double of anything. Rather, it is the exponentiation of a classical Lie algebra double with zero cobracket in line with what we have explained above.

Finally, since the above action of $SL_2(\C)$ on $SU_2^\star$ is quite complicated, it can be helpful to write the latter in a more suitable form as  the upper half of  the two-sheeted hyperboloid in 3+1 Minkowski space. This is also topologically $\R^3$ and comes with its own natural hyperbolic
 metric induced from the inclusion. The group structure  is not manifest in this description, however. 
 To give the change of coordinates we write elements  $x$ of  Minkowski space  as  $2\times 2$ Hermitian matrices, with determinant 1 for the unit hyperboloid. An element $g\in SL_2(\C)$ act on such a matrix  via  $x\mapsto g^\dagger x g$. We identify the unit matrix (the point $(1,0,0,0)$ in usual time-space form) here with the unit matrix of $SU_2^\star$. Our factorisation of $SL_2(\C)$ is exactly into the subgroup $SU_2$ of spatial rotations that leaves this point invariant and the subgroup of boosts which is $SU_2^\star$ and acts by (in the conventions above) right multiplication. Thus a general point of $SU_2^\star$ corresponds to a $2\times 2$ Hermitian matrix in the upper half hyperboloid by
\[  \begin{pmatrix}w &z \\ 0 & w^{-1}\end{pmatrix}\leftrightarrow  \begin{pmatrix}w &z \\ 0 & w^{-1}\end{pmatrix}^\dagger \begin{pmatrix}1 & 0\\ 0 & 1\end{pmatrix}  \begin{pmatrix}w &z \\ 0 & w^{-1}\end{pmatrix}= \begin{pmatrix}w^2 &wz \\ w\bar z & w^{-2}+z\bar z\end{pmatrix}.\]
One can coordinatise $SU_2^\star$ with coordinates of length dimension  in a variety of ways, for example
\[ w=1+\lambda \CX_3,\quad z=\lambda(\CX_1+\imath \CX_2),\quad \CX_3> -{1\over\lambda}.\]
Then the group structure appears as a modified addition law of $\R^3$, cf\cite{Ma:book}. Equipped with a compatible Riemannian metric, hyperbolic space is  a  curved  deformation of  $\R^3$, becoming flat  in the limit $\lambda\to 0$. 
One also has Lie algebra coordinates $x_a$ with matrix $e^{\imath\vec x\cdot \vec \rho}$ for certain matrices $\rho_a$. The exponential map here is a bijection with $\R^3$.

The model has a similar physical interpretation to that of Section~3.3, i.e. quantum particles on a classical background with curvature (due to the presence of a cosmological constant) but uncoupled to the sources. The difference is that the motion is locally described by motion on hyperbolic 3-space with $SL_2(\C)$ transitions between patches.

\subsection{$U(su_2)\rlbicross \C[SU_2^\star]$ -- semidual of  free particle in hyperbolic space (bicrossproduct spacetime)}
 
Next, we apply the semidualisation construction to the preceding  model with spacetime curvature. 
Once again,  this interchanges the role of position and momentum at a Hopf-algebraic level. Hence
space becomes the flat but noncommutative `bicrossproduct spacetime' whose coordinate algebra is the enveloping algebra  $U(su_2^\star)$, i.e. with non-zero brackets
\begin{equation}
\label{kappax}
[x_i,x_3]=\imath\lambda x_i
\end{equation}
for $i=1,2$, where the deformation parameter  $\lambda$  should be interpreted  as $1/m_p$ 
in this model.  Meanwhile, rotations remain unchanged as $SU_2$ or $U(su_2)$ at the Hopf algebra level while
the enveloping algebra of momentum is the commutative algebra of functions on $SU_2^\star$. This is the bicrossproduct euclidean quantum group. Its dual can be viewed as quantizing the bicrossproduct Poisson-Lie group $SU_2\rlbicross su_2$ where $su_2$ is an additive group, with a certain bicrossproduct Poisson-Lie structure \cite{Ma:ista}. The classical group here is once again $E_3$ but with a different Poisson-Lie group structure than in some of the above models.

 To give details, in order to have all quantum groups left-acting, we again flip conventions to a conjugate factorisation $SL_2(\C)=SU_2^\star \cdot SU_2$,  given by
\[ \begin{pmatrix}a&b\\ c&d \end{pmatrix}=\begin{pmatrix}w &0 \\ z & w^{-1}\end{pmatrix}\begin{pmatrix}x & y \\ -\bar y & \bar x\end{pmatrix},\quad |x|^2+|y|^2=1,\quad w>0,\quad x,y,z\in\C,\]
\[ w=\sqrt{|a|^2+|b|^2},\quad x=w^{-1}a,\quad y=w^{-1}b,\quad z=w^{-1}(\bar a c+\bar b d).\]  This implies a Hopf algebra factorisation $U(sl_2)=U(su_2^\star)\dcross U(su_2)$  as a version of the classical cosmological model above. Semidualisation using the $A$-version of the theory (in the terminology  of Section~2.6) then gives a new Hopf algebra $U(su_2)\rlbicross \C[SU_2^\star]$ which acts canonically on $U(su_2^\star)$.  This can be computed explicitly cf.\cite{Ma:ista,Ma:book}
\[ [J_a,J_b]=\imath\eps_{abc}J_c,\quad [P_a,J_3]=\imath\eps_{a3c}P_c,\quad [P_3,J_a]=\imath\eps_{3ab}P_b\]
\[ [P_a,J_b]={\imath\over 2}\eps_{ab3}\left({1-e^{-2\lambda P_3}\over \lambda}-\lambda (P_1^2+P_2^2)\right)+\imath\lambda\eps_{ac3}P_bP_c,\]
giving  a nonlinear action of $su_2$ on the manifold of $SU_2^\star$. This manifold can be naturally identified with    hyperbolic  space, as explained at the end of Section~3.4. Meanwhile, as indicated in the bicrossproduct notation, the coalgebra also has a semidirect form
\[    \Delta J_i=J_i\tens 1+e^{-\lambda P_3}\tens J_i+\lambda P_i\tens J_3,\quad \Delta P_i=P_i\tens 1+e^{-\lambda P_3}\tens P_i\]
for $i=1,2$ and the usual additive ones for $P_3,J_3$. 

The action of this quantum group on the bicrossproduct position algebra $U(su_2^*)$ is 
\[ J_a\la x_b=\imath\eps_{abc}x_c,\quad P_a\la:f(x):=:{\del\over\del x_a}f(x):\]
where $:\ :$ denotes normal ordering of an ordinary polynomial with $x_3$ to the right.

\subsection{$D(U_q(su_2))$ -- quantum gravity with cosmological constant ($q$-hyperbolic spacetime  $B_q[SU_2]$)}

Finally, we can follow the same ideas but now in  quantum gravity with cosmological constant, where there are no classical groups or spaces on either side of the semidualisation. We are actually going to give some different versions algebraically equivalent when $q\ne 1$ by `purely quantum' isomorphisms.
Note that for the quantum group $U_q(su_2)$ we use the standard generators $H,X_\pm$ generators for so that
\begin{equation}
\label{standardq}
 q^{H\over 2}X_\pm q^{-{H\over 2}}=q^{\pm 1}X_\pm,\; [X_+,X_-]={q^H-q^{-H}\over q-q^{-1}},
 \end{equation}
as well as 
\[\Delta q^{\pm\frac{H}{2}}= q^{\pm\frac{H}{2}}\tens  q^{\pm\frac{H}{2}},
\quad
\Delta X_\pm=q^{-{H\over 2}}\tens X_\pm+ X_\pm\tens q^{{H\over 2}}.
\]
The real form here is defined by $H^*=H$ and $X_\pm^*=X_\mp$ at least for real $q$ (the root of unity case is more subtle). For its dual $\C_q[SU_2]$ we use a matrix of generators $t^i{}_j=\begin{pmatrix}a&b\\ c&d\end{pmatrix}$, with its usual relations 
\[ ba=qab,\; bc=cb, \; bd=q^{-1}db,\;ca=qac,\;
 cd=q^{-1}dc,\; da=ad+(q-q^{-1})bc\]
and matrix form of coproduct. The real form is given by $a^*=d, b^*=-q^{-1}c$ for $q$ real.

For our first version in Figure~1, the form suggested by the classical geometry is the quantum double viewed as
\[ U_q(so_{1,3})=U_q(su_2)\dcross U_q(su_2^\star),\]
where $U_q(su_2^\star)\isom \C_q[SU_2]^{op}$ with new generators $\xi,x$ and $y$  defined by
\begin{equation}\label{qSU2xy}\begin{pmatrix}a&b\\ c&d\end{pmatrix}=\begin{pmatrix}q^\xi & \lambda y\\ \lambda x& q^{-\xi}(1+q\lambda^2xy)\end{pmatrix},\quad \lambda=q^{-1}-q,\end{equation}
and relations and coproduct that the reader can translate. For example, the relations here are 
\begin{equation}
\label{anagain}
[\xi,x]=x, \quad [\xi,y]=y, \quad [x,y]=0,
\end{equation}
 so as an algebra it is in fact $U(su_2^\star)$, undeformed. This is the `purely quantum isomorphism' on the lower left in Figure~1, valid for $q\ne 1$. Note that in this model the small deformation parameter $\lambda\approx 2/(m_p l_c)$ is, like $q$, dimensionless. The quantum double in this form is the dual of the quantum group quantising $su_2\dcross su_2^\star$ with its classical double Poisson Lie group structure. There is a canonical action on $U_q(su_2)^{cop}=U_{q^{-1}}(su_2)$ with generators $h,x_\pm$, say, (to distinguish from the previous ones) and relations with inverted $q$. This  could serve  as a definition of $\C_q[SU_2^\star]$ as a noncommutative space with generators $w$ and $z$ defined via
\[  \begin{pmatrix}w &z \\ 0 & w^{-1}\end{pmatrix}= \begin{pmatrix}q^{h\over 2} &q^{-{1\over 2}}\lambda x_-\\ 0 & q^{-{h\over 2}}\end{pmatrix},\]
a matrix form of coalgebra and relations that the reader can translate from those of $U_q(su_2)$. One needs the complex conjugate as an additional generator $z^*$ of $\C_q[SU_2^\star]$ to complete this to a $*$-algebra along with $w^*=w$ as a real generator. This version of the model is a $q$-deformation of the free particle on hyperbolic spacetime (the middle left model of Figure~1, Section~3.4), with $q$-deformation the introduction of finite $m_p$ or the  `switching on' of mutual gravitational interaction via the Newton constant $G$. 

Next, as  in the classical case, it is  natural to define this $q$-hyperbolic space  as the  unit mass-hyperboloid of $q$-Minkowski space. The necessary $q$-Minkowski space is defined as the coordinate algebra $B_q[M_2]$  of the space of  $2\times 2$ braided Hermitian matrices \cite{Ma:exa,Ma:book}
\[ \beta\alpha=q^2\alpha\beta,\quad \gamma\alpha=q^{-2}\alpha\gamma,\quad \delta\alpha=\alpha\delta,\]
\[ [\beta,\gamma]=(1-q^{-2})\alpha(\delta-\alpha),\quad [\delta,\beta]=(1-q^{-2})\alpha\beta,\quad
[\gamma,\delta]=(1-q^{-2})\gamma\alpha,\]
\[\underline{ \Delta}\begin{pmatrix}\alpha&\beta\\ \gamma&\delta\end{pmatrix}=\begin{pmatrix}\alpha&\beta\\ \gamma&\delta\end{pmatrix}\tens \begin{pmatrix}\alpha&\beta\\ \gamma&\delta\end{pmatrix},\]
\[ \underline{\eps}\begin{pmatrix}\alpha&\beta\\ \gamma&\delta\end{pmatrix}=\begin{pmatrix}1&0\\ 0&1\end{pmatrix},\quad \begin{pmatrix}\alpha&\beta\\ \gamma&\delta\end{pmatrix}^*=\begin{pmatrix}\alpha&\gamma\\ \beta&\delta\end{pmatrix}.\]
The coproduct here extends to products with braid statistics, much as for super-matrices but with bose-fermi statistics replaced by a braiding matrix. If we quotient by the braided-determinant relation
$\alpha\delta-q^2\gamma\beta=1$
we have the unit hyperboloid in $q$-Minkowski space, which is the coordinate algebra of the braided group $B_q[SU_2]$. The $q$-determinant otherwise defines a $q$-metric.  When $q\ne 1$ this algebra is more or less isomorphic to $U_q(su_2)$ as required by means of 
 the `quantum Killing form', as 
\[ \begin{pmatrix}\alpha &\beta\\ \gamma &\delta\end{pmatrix}= \begin{pmatrix}w &z \\ 0 & w^{-1}\end{pmatrix}^* \begin{pmatrix}w &z \\ 0 & w^{-1}\end{pmatrix}=\begin{pmatrix}q^h & q^{-1\over 2}\lambda q^{h\over 2}x_-\\ q^{-1\over 2}\lambda x_+q^{h\over 2}& q^{-h}+q^{-1}\lambda^2x_+x_-\end{pmatrix}\]
in terms of our previous identification. This quantum Killing form can also be viewed more categorically as essentially an isomorphism between the braided enveloping algebra $BU_q(su_2)$ (which has the same algebra as $U_q(su_2)$) and its dual which is the braided function algebra $B_q[SU_2]$. 

For our second version of $D(U_q(su_2))$ we come from the quantum double construction rather than the classical version.  So we work with $D(U_q(su_2))=U_q(su_2)\bowtie \C_q[SU_2]^{op}$ acting likewise on $U_q(su_2)^{cop}$ viewed as $\C_q[SU_2^\star]$ or by preference as $B_q[SU_2]$.  Moreover, it turns out to be very natural to replace  $\C_q[SU_2]^{op}$ in the quantum double by another copy of $B_q[SU_2]$ with matrix generators $u^i{}_j$, say. Then one finds
\[ D(U_q(su_2))\cong U_q(su_2)\rbiprod B_q[SU_2],\]
which is then a semidirect product as an algebra and as a coalgebra, called the `bosonisation' of $B_q[SU_2]$ \cite{Ma:book}. Here $U_q(su_2)$ acts on $B_q[SU_2]$ both as spacetime and as rotations by the quantum coadjoint action.  This form  of the quantum double expresses the model as a $q$-deformation of quantum gravity without cosmological constant in Section~3.2, i.e. as purely introducing the cosmological constant.

Finally, using this braided theory we are able better to understand our first version, as a third formulation of the quantum double
\[ U_q(so_{1,3})= U_q(su_2)\codcross U_q(su_2),\]
which as an algebra is the tensor product one. This describes $U_q(so_{1,3})$ as a complexification of $U_q(su_2)$ and a further `twisting' of the coproduct. This form of the quantum double follows from the  $U_q(su_2)\rbiprod BU_q(su_2)$  form (using the quantum Killing form isomorphism above) and the fact that the semidirect product by the quantum adjoint action used for the algebra structure can then be unravelled to a tensor product. This explains our two points of view of the model as  shown on left side of the lower block in Figure~1. They are isomorphic provided $q\ne 1$, a `purely quantum' phenomenon.

\subsection{$U_q(so_4)$ -- semidual of quantum gravity with cosmological constant ($\C_q[SU_2]$ spacetime) and self-duality}

The semidual of the preceding quantum double model has quantum group $U_q(su_2)\tens U_q(su_2)^{cop}=U_q(so_4)$ acting on the $q$-deformed space $\C_q[SU_2]^{op}$. The action here is by left and right differentials, i.e. by the coproduct of $\C_q[SU_2]$ viewed as a left or right coaction and evaluated against the two copies of $U_q(su_2)$. This version of the model exactly $q$-deforms the semidual of quantum gravity without cosmological constant based on $\widetilde{SO_4}$ acting on $SU_2$, i.e. it $q$-deforms the free particle on $SU_2$ with cosmological constant (the upper right of Figure~1) with $q$-deformation 
introducing  mutual gravitational interactions via finite $m_p$ or non-zero Newton constant $G$.

\begin{thm} For generic $q\ne 1$, or for the reduced theory at $q$ a root of unity, quantum gravity with cosmological constant as above is self-dual up to an algebraic equivalence under semidualisation. The algebraic equivalence is given by a quantum Wick rotation \cite{Ma:euc} or `transmutation' from $\C_q[SU_2]^{op}$ to $B_q[SU_2]$ as spacetime algebra and a Drinfeld twist from $U_q(su_2)\codcross U_q(su_2)$ to $U_q(su_2)\tens U_q(su_2)^{cop}$ as $q$-isometry group.
\end{thm}

The Drinfeld twist needed is the composition of two; one to convert $U_q(su_2)^{cop}$ to $U_q(su_2)$ and the second to convert $U_q(su_2)\tens U_q(su_2)$ over to $U_q(su_2)\codcross U_q(su_2)$. The Drinfeld twist here conjugates the coproduct by a Hopf-cocycle. Its key feature is that {\em it does not change the category of modules up to a formal equivalence}. More precisely, since in this form the algebras of the two quantum groups are the same, their category of modules have the same objects. Tensor products of two modules depend on the coproduct and these are related by a twisting cocycle obtained from the braiding or `universal $R$-matrix' of $U_q(su_2)$ (so the tensor products are nontrivially isomorphic by this cocycle). Details were developed by one of the authors in the early 1990s and are in \cite{Ma:book} and elsewhere. To do this rigorously, however, one has to look at the the convergence of powerseries or work not over $\C$ but over the ring of formal power-series in the deformation parameter. The reader can also say quite rightly that the categories of modules of $U(so_{1,3})$ and $U(so_4)$ are quite different and cannot possibly coincide. Indeed, the only difference in the classical case is the $*$-structure or unitarity constraint. However, in the $q$-deformed theory $U_q(so_{1,3})$ and $U_q(so_4)$ are different even as Hopf algebras  and it is at this algebraic level that we have the equivalence (i.e., not respecting the $*$-structures of the quantum groups, which are not equivalent). Also, in physical terms the situation is actually {\em more} precise when $\Lambda>0$  i.e. when $q$ is a complex number of modulus 1,  and we look at the truncated theory at $q$ a root of unity.  In this case we must use finite `reduced' versions of all our algebras and have exact isomorphisms. Some theory of $\C_q[SU_2]$ at $q$ a root of unity is in \cite{GomMa:coh}. 

The specific twists here also have a deep braided category interpretation which is the origin of the term `transmutation'. This theory converts ordinary quantum groups such as $\C_q[SU_2]$ into braided ones such as $B_q[SU_2]$ but in such as way that all of the theory has braided parallels. In particular, there is also a braided version $BU_q(su_2)$ of $U_q(su_2)$ and the (essentially) isormorphism $B_q[SU_2]\cong BU_q(su_2)$ has a categorical origin as braided-selfduality of such `factorisable' quantum groups. Because of it, the braided Fourier transform becomes an operator $B_q[SU_2] \to B_q[SU_2]$ which, together with left multiplication by the ribbon element generates a representation of the mapping class group $PSL(2,\Z)$\cite{LyuMa:bra}. This representation is at the heart of the three-manifold invariant corresponding to the quantum group $U_q(su_2)$. The same applied to $D(U_q(su_2))$ is at the heart of the Turaev-Viro invariant, i.e. of the solution of this part of 3d quantum gravity with cosmological constant.  Moreover, because the quantum gravity theory with point sources is controlled essentially by attaching representations at the marked points of the Riemann surface as explained in Section~2, the semidual $U_q(so_4)$ theory has in some sense the same physical content up to the mentioned (but non-trivial) isomorphisms. 

Finally, using the dual of the `purely quantum isomorphism' (\ref{qSU2xy}) we arrive at the other version on the lower right of Figure~1 with quantum isometry group $U_q(su_2)\lrbicross \C_q[SU_2^\star]$, isomorphic when $q\ne 1$. We arrive this time at the $q$-deformation of the bicrossproduct model of Section~3.5, so $q$-deformation
 is now interpreted once again as introducing the cosmological constant.

\subsection{Degenerations between the models}

As indicated in Figure~1 the various models as well as being related by semidualisation horizontally are related vertically by `deformation' going downwards or, going the other way, by degeneration. In this subsection we explain these degeneration maps between the models. The key observation is that the $q$-deformed models in Section~3.6 and~3.7 (the bottom of Figure~1) have only one parameter $q$ which is dimensionless. However, the limit $q\to 1$ can be taking in different ways according as how the generators are also scaled and this gives the various degenerations.

We start at the bottom left of Figure~1, the dimensionless model of Section~3.6. For the rotational part of the isometry quantum group there is no problem and we set  $U_q(su_2)\mapsto U(su_2)$ as $q\to 1$. However, $U_q(su_2)$ is {\em also} the quantum spacetime algebra in the model at the bottom left of Figure~1, and here we must be more careful to write
\[ H=2 m_p x_3,\quad x_\pm= m_p(x_1\pm i x_2)\]
and then take the limit $l_c\to \infty$ in the relations \eqref{standardq} of $U_q(su_2)$. We then obtain the spin model spacetime of Section~3.2 (the upper left of Figure~1) with relations (\ref{ncspace}) for the $x_a$.

On the other hand we can make use of the `purely quantum isomorphism' (\ref{qSU2xy})  and set
\[ \xi=-i l_cP_3,\quad x=l_c(P_1+iP_2),\quad y=-x^* \]
and then take the limit $m_p\to \infty$. We then obtain the momentum sector of the classical hyberbolic spacetime model of Section~3.4 (left middle of Figure~1): the relations \eqref{anagain}
become  the relations for the $P_a$ in \eqref{jphyp}

Similarly on the right hand side of Figure~1 starting at the bottom in the dimensionless $q$-deformed theory of Section~3.7 we can set $q\to 1$  {\em after} identifying the $q$-bicrossproduct as $U_q(su_2)\tens U_q(su_2)^{cop}$. We can set $q\to 1$ here and for the spacetime $C_q[SU_2]^{op}$ to obtain the particle on a 3-sphere in Section~3.3 (upper right in Figure~1). Finally, on the other side of the `purely quantum isomorphism' (\ref{qSU2xy}) we can write
\[ \xi=i m_p x_3 ,\quad x= m_p (x_1+ix_2), \quad y=-x^*,\]
and then take the limit $l_c\to \infty$ to obtain the bicrossproduct spacetime model of Section~3.5 (right middle of Figure~1): the relations \eqref{anagain} turn into  the spacetime algebra \eqref{kappax} for the $x_a$.

We have described here the degenerations at the level of spacetime and isometry algebras. The same applies when one looks deeper into the noncommutative differential geometry of the models. For example, the standard 4D bicovariant differential calculus on $C_q[SU_2]^{op}$ at bottom right degenerates to a 4D quantum-isometry covariant differential calculus on the bicrossproduct spacetime. This will be given in detail elsewhere. The final degenerations to the $E_3$ model of Section~3.1 at the top of the figure are obvious as the remaining $m_p$ or $l_c$ parameter is set to infinity.

\section{Physics of semiduality from spin spacetime to classical $SU_2$}

So far we have been describing our models in terms of the algebraic structure of isometry (quantum) groups  and their relation by semidualisation.   In this section we now look in detail at the physics  in the sense of the irreducible representations in these models, concentrating on the upper part of our overview in Figure~1.   Our main motivation is to understand the physical interpretation of semiduality, using  the strategy outlined in the introduction:
by studying  irreps of   the  isometry (quantum)  groups  and their  semiduals, and using (quantum) Fourier transforms to switch from  momentum- to position representation within one model we are   able  to   realise representations of mutually semidual (and generally non-equivalent) models in terms of functions on the {\em same} (possibly noncommutative) space.  An additional motivation for studying irreps and their Fourier transforms comes from the role of quantum doubles in  the construction of the Hilbert space \eqref{Hilbertius} of 3d quantum gravity,  where the irreps  represent  the one-particle contributions.  While the literature on 3d quantum gravity has focussed on the momentum space picture of those irreps, the (noncommutative) position picture may provide insights into the interpretation of 3d quantum gravity in terms of noncommutative geometry.

We recall that  semiduality maps the euclidean group  $E_3$ to itself
 (but exchanges momenta and positions), whereas the quantum double $D(U(su_2))$ (spin model)
is mapped to the universal enveloping algebra of $so_4$ 
($S^3$ spacetime model). The case of $E_3$ is thus exceptional in that semiduality and Fourier transform coincide.  We give 
give the irreps of $E_3$ both in terms of vector-valued functions on
momentum space satisfying a (multiplicative) spin constraint 
of in terms of vector-valued functions on position space satisfying 
 a Dirac-type linear first order wave equation; the two pictures are
 related via standard Fourier transform. This case is of course the well-known Wigner construction but we present it in a geometrical form that is suitable for deformation. For  $D(U(su_2))$ or more precisely  $D(SU_2)$ in a global formulation, 
the irreps are given precisely as a $1/m_p$-deformation of the $E_3$ picture, both in terms of  vector-valued functions on
curved momentum space $SU_2$ satisfying a (multiplicative)
group-valued spin constraint cf \cite{schroers} and in terms of noncommutative wave equations
on fuzzy $\R^3$ as in \cite{BatMa:non}.  
For $so_4$ the irreps are given in terms of vector-valued functions on
curved position space $SU_2$ satisfying a  linear first order 
differential equation,  and, after quantum  Fourier transform,  in terms of vector-valued functions on fuzzy momentum space  satisfying an algebraic constraint. We show that this, too, is a deformation of the $E_3$, this time recovered as $l_c\rightarrow \infty$. Thus we obtain a precise
dictionary between  the physical pictures in the two non-trivial models. They  are not equivalent, but are both deformations of the same pictures in the $E_3$ case. 

\subsection{Representations of $E_3$}

We first recall some standard facts and notations for $su_2$ and its representations.
We introduce a set of Hermitian generators $t_a$ satisfying the standard commutation relations
\[ [t_a,t_b]=\imath\eps_{abc}t_c\]
and given explicitly via $t_a=\sigma_a/2$ in terms of Pauli matrices. We will denote the $(2s+1)$ dimensional irreducible representation of 
the Lie algebra $su_2$ by $\rho^s$, where $s\in {1\over 2}(\N\cup\{0\})$. 
This has a lowest weight vector which
we will denote by $|s,-s\>$, where $\rho^s(t_3)|s,-s\>=-s|s,-s\>$ in our conventions. For $s=1$ it will be convenient to consider the Cartesian basis, where 
\bea
\label{cart}
\rho^1(t_a)_{bc}=-\imath\epsilon_{abc},
\eea
and for $s=1/2$ it will be convenient to use the defining Pauli matrix representation $\rho^{1/2}(t_a)=t_a=\frac 1 2 \sigma_a$. 
We will also use the $t_a$ basis to identify $su_2$ with $\R^3$. However, all of our constructions are basis independent. 

The euclidean group $E_3=SU_2 \ltimes \R^3$ was covered in Section~\ref{e3sect} and we use the notations from there. In particular,
recall that the translation part is identified with $su_2^*$, 
with generators denoted $P_a$ so that  a finite translation is written as $a=-\imath a_bP_b$.
According to the standard theory, irreps of $E_3$ are labelled by $SU_2$ orbits in momentum space $(su_2^*)^*$ together
with  irreps of associated centralisers. Since  $(su_2^*)^*=su_2$, momentum space is $su_2$ and 
we could use the basis $\{t_a\}$, but we need to be careful about normalisation. As explained in Section~\ref{e3sect}
the  dual basis $\{P_a^*\}$ may have a different normalisation from   that of  $\{t_a\}$, which is fixed by the commutation
relations, so we should allow
\bea
\label{ptrel}
P_a^* = -\lambda t_a,
 \eea
where $\lambda$ is an arbitrary constant of dimension 
inverse mass.
Thus  we    view $su_2$ as momentum space and denote elements as  $p$, which we expand
as 
\bea
p=\imath p_a P_a^* = -\imath \lambda  p_a t_a
\eea
if we wish to use an $\R^3$ notation. We should stress that the parameter $\lambda$ only enters the discussion
because we choose to work with the basis $\{t_a\}$ of momentum space; if we carried out the analysis entirely in terms
of the  basis $\{P_a^*\}$ this parameter would not be required. 

The irreducible representations of $E_3$ are then  labelled by adjoint 
$SU_2$ orbits i.e. by  two-spheres $S^2_\m=\{v\lambda \m  t_3
v^{-1}\,| \, v\in SU_2\}$ in momentum space  
  and irreducible unitary
representations $\Pi_s$  of associated  stabilisers
$N_\m=\{ g\in SU_2| g\lambda \m  t_3  g^{-1}=\lambda \m  t_3\}$.
Clearly $N_0 \simeq SU_2$ and $N_{\m}\simeq U(1)$
for all other values of $\m$ and $s\in \frac 1 2 (\N\cup\{0\})$. 
The parameters $m$ and $s$ are interpreted as (euclidean) mass and spin of a particle.
In the generic case the carrier
spaces for the  irreducible representations are
\bea
\label{eurep}
\quad &V_{\m s}=&\{ \psi:SU_2 \rightarrow \CC \;|\; \psi(ve^{\alpha\imath t_3}) e^{is \alpha} 
\psi(v),\ \forall \alpha \in [0,4\pi) ,\ \forall v \in SU_2\},
\eea
whose elements also arise as sections of Dirac monopole bundles, and we therefore refer to them as
monopole sections.
An element $(g,a) \in E_3$ acts on a monopole section  via 
\bea
\label{e3rep}
\pi_{\m s}((g,a))\psi (v)= \exp({\imath\m a\left( \Ad_{ g^{-1} v } (\imath P_3^*)\right) })\psi (g^{-1}v).
\eea
If we introduce  the $su_2$ element
\bea
\label{pdef}
p=\imath m vP^*_3v^{-1},
\eea
the phase here could be written as 
\[
\exp({\imath  \vec a\cdot\Ad_{g^{-1}}(\vec p)})
 \]
 when both $a$ and 
$p$ are expanded in the mutually dual  bases  $\{-\imath P_a\}$ and 
$\{\imath P_a^*\}$.  For $\m=0$ the centraliser representations are $SU_2$ representations. In the resulting finite-dimensional 
representations of $E_3$, the translations act trivially.
We  are not interested in  the finite dimensional irreducible representations 
in the following.

Given $\psi \in V_{\m s }$ define the map
\bea
\label{step1}
\tilde\phi: S_\m^2 \rightarrow \CC^{2s+1},
\eea
where $S^2_\m$ is the  two-sphere in $su_2$ of radius $\lambda \m$, via
\bea
\label{phidef}
\tilde\phi(p)=\psi(v)\rho^s(v)|s,-s\>,
\eea
where $p$ is related to $v$ via \eqref{pdef}.
Clearly 
\[ \rho^s(ve^{\alpha\imath t_3})|s,-s\>=\rho^s(v)\rho^s(e^{\alpha \imath t_3})|s,-s\>=\rho^s(v)e^{-\imath\alpha s}|s,-s\>\]
which cancels the phase picked up by $\psi$ under the  right-multiplication by 
$e^{\alpha\imath t_3}$. Hence 
 $\tilde\phi$ only depends on $p\in S^2_\m$ even though both $\rho^s(v)$ 
and $\psi$ depend on $v$. 

The map $\tilde\phi$ defined in \eqref{phidef} satisfies the constraint
\bea
\label{momconstraint}
(\rho^s(t_a) p_a +\m s)\tilde\phi=0.
\eea
To see this,  write \eqref{pdef} as $p_at_a=vmt_3v^{-1}$ so that  
\bea
\rho^s(t_a)p_a\tilde\phi(p)&=&\rho^s(v\m t_3 v^{-1})\rho^s(v)\psi(v)|s,-s\>
\nonumber \\
&=&\psi(v)\rho^s(v)\m (-s) |s,-s\>\nonumber \\&=&-\m s \tilde\phi(p), \nonumber 
\eea
as required. Conversely, any map $\tilde\phi:S^2_{\m}\rightarrow \CC^{2s+1}$
satisfying this  constraint can be written in 
the form \eqref{phidef}
with $\psi\in V_{\m s}$.  Thus  the field $\tilde\phi(p)$ is the monopole section corresponding to $\psi$ but written `down stairs' on the base $S^2_\m$ of the monopole bundle as a function with values in a one-dimensional vector space within $\C^{2s+1}$ that varies as we move about on the base, in other words as  an element in a rank 1 projective module. There is an associated  projection matrix
at every point $p\in S_m$:
\bea
\label{projector} 
e(p)=\rho^s(v)|s,-s\>\<s,-s|\rho^s(v^{-1}), 
\eea
with \eqref{pdef} assumed,  which projects any $\tilde\phi$ to a solution of our constraint, i.e. down to the irreducible representation. Notice that for $s=1/2$ we have 
\bea 
\label{epspinhalf}
e(p)={1\over 2}+{t_ap_a\over\m}, 
\eea
while for other spins the relationship is more complicated.

To obtain a unified description of all (infinite-dimensional)
irreducible
representations we consider the union
\[
\bigcup_{\m \in \RR^+} \,\,S^2_\m \simeq \RR^3 \setminus \{0\} 
\]
and use the carrier space 
\[
W_{ s}=\{\tilde\phi: \RR^3 \setminus \{0\}  \rightarrow \CC^{2s+1}\}
\]
as a starting point for the representation
theory of $E_3$. The subspaces 
\[
W_{\m s}=\{\tilde\phi: \RR^3 \setminus \{0\}  \rightarrow \CC^{2s+1}|
(\rho^s(t_a) p_a +\m s)\tilde\phi=0\}
\]
obtained by imposing the constraint  are representation of $E_3$. In order to obtain an irrep as before we may still need
to impose an additional constraint
\[ (p^2-\m^2)\tilde\phi=0\]
although for spins 1/2, 1 this holds automatically. An element $(g,a) \in E_3$ acts  via 
\[
\pi_{\m s}((g,a))\tilde\phi (p)= e^{\imath a(\Ad_{g^{-1}}p)}\,\rho^s(g)\,\tilde\phi(\Ad_{g^{-1}}p),
\]
which commutes with the constraint \eqref{momconstraint}, as required. The angle in the phase here is 
again $\vec a\cdot\Ad_{g^{-1}}\vec p$ in our chosen bases.

The advantage of working with the map $\tilde\phi$ in this way is that it
is defined on a linear space. We can Fourier transform back to a field
\[
\phi(x)=\frac{1}{\sqrt{(2\pi)^3}}\int d^3p
\,\, e^{\imath\vec x\cdot\vec p}\tilde \phi(p),
\]
which turns the constraint \eqref{momconstraint} into the 
first order differential equation
\bea
\label{posconstraint}
(\imath\rho^s(t_a) \partial_a -\m s) \phi=0.
\eea
For $s=\frac 1 2 $ this is the Dirac equation
\bea
\label{dirac}
(\imath\sigma_a \partial_a -\m)\phi= 0. 
\eea
Applying the adjoint Dirac operator $i\sigma_a \partial_a +\m$
we deduce
\[
(\Delta+\m^2)\phi =0.
\]
For $s=1$ the equation \eqref{posconstraint} takes the form
\bea
\label{massive}
\nabla \times \phi = -\m \phi,
\eea
where we used the Cartesian representation \eqref{cart}.
 Computing the divergence on both sides 
we deduce $\nabla\cd\phi=0$ and therefore, upon applying $\nabla
\times$
to both sides of \eqref{massive}, 
\[
(\Delta+\m^2)\phi =0.
\]

To sum up, we obtain irreducible representations of $E_3$  on the
space
of $\CC^{2s+1}$ valued ``wave functions'' satisfying a first order
 equation, which generalises the Dirac equation
\[
W_{\m s}=\{ \phi: \RR^3 \rightarrow \CC^{2s+1}|
\ (\imath \rho^s(t_a)\partial_a -\m s) \phi=0\},
\]
at least for spin $1/2$ and $1$. For higher spins one may need to supplement with
the usual wave equation  $(\Delta+\m^2)\phi$ as for scalar fields.  An element $(g,a) \in E_3$ acts
on a wavefunction  via 
\[
\pi_{\m s}((g,a))\phi \,\,(\vec x)= \rho^s(g)\, \phi(\Ad_{g^{-1}}(\vec x)-\vec a).
\]
The infinitesimal generators $P_a$ and $J_a$ of translations and rotations act as 
\bea
\label{e3gen}
P_a=-\imath\frac{\partial}{\partial x_a}, 
\qquad J_a=-\imath\epsilon_{abc} x_b\frac{\partial}{\partial
  x_c} + \rho^s(t_a),
\eea
so that $\vec P\cdot\vec J=-\imath \rho^s(t_a){\partial_a} $ is the
Casimir used in the definition of $W_{\m s}$.

\subsection{Representations of the quantum double $D(SU_2)$}

We now look similarly at the particle states in the quantum double  `spin model' related to 3d quantum gravity without cosmological constant. We will view the quantum double here as a deformation of $E_3$ \cite{schroers, BatMa:non}
as we explained in Section~3.2, with a parameter $\lambda=8\pi/m_p$ in the quantum gravity application. Note that, with this choice for $\lambda$, the relation  \eqref{ptrel} between  rotation and dual translation generators is 
the identification  \eqref{pjrel}
of  $J_a$ with $P_a^*$ in terms of the non-degenerate symmetric from used
in the Chern-Simons action for 3d gravity. As we shall see, the identification of $P_a^*$ with $J_a$ (or $t_a$), which 
was optional in the discussion of $E_3$ representations, is essential in the following discussion 
of  quantum double representations. Our treatment is  fully analogous to the one of $E_3$, including a physical interpretation as particles of some kind with mass and spin. 

We start with some remarks about the relevant quantum double.
Indeed, the required quantum double of a compact Lie group $G$  has been studied  in
various publications and can be defined as a particular Hopf $C^*$-algebra. However, its formulation as such is quite technical and in practice one can take either a $*$-algebraic approach in terms of generators and relations, much as in physics one can work at the Lie algebra level in practice. Thus,  $D(U(\g))=U(\g)\rcross \C[G]$ where $U(\g)$ denotes the enveloping algebra of the
Lie algebra of `rotations' (in our application) and $\C[G]$ an algebra of coordinates in momentum space $G$. The semidirect product is by the right adjoint action and in the case of $SU_2$ the required structure was given in Section~3.2 as derived in \cite{BatMa:non}. Note, however, that group elements do not themselves lie in $U(\g)$ but in a completion, i.e. have to be approximated.

The more technical $C^*$ approach makes use of a cross product $C^*(G)\rcross C(G)$ of the group $C^*$-algebra and the $C^*$-algebra of continuous functions on $G$. The former is defined first by a convolution product of functions of compact support and then completed. A closely related approach \cite{KM}  is to start with continuous
functions on $G\times G$ with convolution on the first factor (note that we exchange the roles
played by the two copies of $G$ in order to match our
conventions for the semidirect product group $E_3$).  In these approaches one obtains eventually a Hopf $C^*$-algebra $D(G)$ but one still has to approximate the actual elements of the   `rotation group' copy of $G$ since these would appear 
as $\delta$-functions in the convolution algebra.  If we  allow these for purposes of
writing simple formulae, we have   multiplication $\bullet$, identity 1,
co-multiplication $\Delta$,
co-unit $\epsilon$, antipode $S$ and involution ${}^*$ via
\bea
(F_1\bullet F_2)(g,u)&:=&\int_G F_1(z,zuz^{-1})\,F_2(z^{-1}g,u)\,dz,  \nonumber 
 \\
1(g,u)&:=&\delta_e(g),  \nonumber \\
(\Delta F)(g_1,u_1;g_2,u_2)&:=&F(g_1,u_1u_2)\,\delta_{g_1}(g_2). \nonumber \\
\epsilon(F)&:=&\int_G F(g,e)\,dg,\nonumber  \\
(S F)(g,u)&:=&F(g^{-1},g^{-1}u^{-1}g), \nonumber \\
F^*(g,u)&:=&\overline{F(g^{-1},g^{-1}ug)}, \nonumber 
\eea
or, entirely in terms of $\delta$-functions,
\bea
(\delta_{g_1}\otimes  f_1)\bullet (\delta_{g_2}\otimes f_2)
&=& \delta_{g_1g_2}\otimes f_1(g_2(\ )g_2^{-1})f_2 \nonumber \\
\Delta(\delta_g\otimes f)(g_1,u_1;g_2,u_2)&=&\delta_g(g_1)
\delta_{g}(g_2) f(u_1u_2)\nonumber \\
\epsilon(\delta_g\otimes f)&=&f(e)\nonumber \\
S (\delta_g\otimes f)&=&\delta_{g^{-1}} \otimes f(g^{-1}()^{-1}g), \nonumber \\
(\delta_g\otimes f)^*&=&\delta_{g^{-1}} \otimes f^*(g^{-1}()g). \nonumber 
\eea
In the following we  will  use  both the algebraic and the group convolution formulations. In the latter  form it is less easy to take the limit to $E_3$ but see \cite{schroers}.

The momentum space is now the curved space $S^3=SU_2$ with `translation Hopf algebra'   given by functions $C(SU_2)$. It acts on another copy of $C(SU_2)$, functions on momentum space, by pointwise multiplication. In a suitable formulation,   the irreducible representations of $D(SU_2)$ 
are labelled by the
  $SU_2$-conjugacy classes $C_\m=\{v e^{\imath\m\lambda   t_3}
v^{-1}\,| \, v\in SU_2\}$
in the momentum space $SU_2$  and irreducible unitary
representations $\Pi_s$  of associated  stabilisers
$N_\m=\{ g\in SU_2| g e^{\imath\m\lambda  t_3} g^{-1}=e^{\imath\m\lambda  t_3}\}$ \cite{KM}.
Note that $C_0=\{1\}$ and $C_{2\pi/\lambda}=\{-1\}$ and that all the 
other conjugacy classes are isomorphic to two-spheres in the Lie algebra
coordinate system, namely $|\vec p|=\m$.
Clearly $N_0\simeq N_{{2\pi}/{\lambda}} \simeq SU_2$ and $N_{\m}\simeq U(1)$
for generic values of $\m$. In the generic case the carrier
spaces for the  irreducible representations are
\bea
\label{doublerep}
&&V_{\m s}= \{ \psi:SU_2 \rightarrow \CC \;|\; \psi(ve^{\imath \alpha t_3}) e^{\imath s \alpha} 
\psi(v),\ \forall \alpha \in [0,{4\pi}), v \in SU_2\}.
\eea
These are the same spaces of monopole sections as before for $E_3$. An element $F\in D(SU_2)$ acts  via 
\[
\Pi_{\m s}(F)\psi (v)= \int dg \, F(g,g^{-1}v e^{\imath\m\lambda t_3} v^{-1}g)\psi(g^{-1}v).
\]
The  singular elements have the simple action 
\[
\Pi_{\m s}(\delta_g\otimes f)\psi (v)= f(g^{-1}v e^{\imath\m\lambda t_3}v^{-1}g) 
\psi(g^{-1}v).
\]

As for the $E_3$ we can alternatively use  carrier spaces which are 
spaces of vector-valued functions 
satisfying a constraint. Again we switch from the function $\psi \in V_{\m s}$ 
to the vector-valued function defined as in \eqref{step1} by
\bea
\label{dsurep}
 \tilde\phi(u)=\rho^s(v)\psi(v)|s,-s\>,
\eea
where now $u=ve^{\imath\m\lambda t_3}v^{-1}\in C_\m$. They are spaces of sections
of a monopole bundle over $C_\m$ with projection
\[ e(u)=\rho^s(v)|s,-s\>\<s,-s|\rho^s(v^{-1})\]
as before but now with the two-sphere viewed as a conjugacy class $C_\m\subset SU_2$ rather than as
an orbit in $su_2$. The functions \eqref{dsurep} satisfy the group-valued analogue 
 of the constraint \eqref{momconstraint},
\bea
\label{groupconstraint}
\rho^s(u)\tilde\phi(u)=e^{-\imath\m\lambda s} \tilde\phi(u),
\eea
as one can check by an analogous calculation to the one carried out after \eqref{momconstraint}.

For a unified description we now   foliate $S^3=SU_2$ as 
\[
\bigcup_{\m \in (0, 2\pi/\lambda )} \,\,S^2_\m \simeq SU_2 \setminus \{1,-1\}. 
\]
Geometrically, $SU_2 \setminus \{1,-1\}$ is the 3-sphere without north and 
south pole, which we denote  $S^3_{NS}$.
We define the space 
\bea
\label{groupfcts}
W^1_{s}=\{\tilde\phi: S^3_{NS}\rightarrow \CC^{2s+1} \}, 
\eea
and impose 
a group-valued
 constraint \eqref{groupconstraint}.
Then we obtain  representations of $D(SU_2)$ on the 
spaces
\bea
\label{groupirreps}
W_{\m s}=\{\tilde\phi: S^3_{NS}\rightarrow \CC^{2s+1}|
\rho^s(u)\tilde\phi(u)=e^{-\imath\m\lambda s} \tilde\phi(u)\}, 
\eea
essentially as before, while to obtain an irrep we may still have to impose a constraint that $\tilde\phi$ has support on $C_\m$ (we will give this in a different coordinate system shortly). For spins $1/2$ and $1$ this is automatic. The action of $D(SU_2)$ is most easily expressed in terms
of the singular elements:
\[
\Pi_{\m s}(\delta_g\otimes f)\tilde\phi (u)= f(g^{-1}ug) \rho(g) \tilde\phi(g^{-1}ug).
\]

In the case of the euclidean group we were able to apply a Fourier
transform to obtain irreducible representations in terms of 
functions obeying a differential equation. We can do just the same in the nonabelian case provided we use the modern tools of quantum group Fourier transform\cite{Ma:ista,BatMa:non,Ma:spo,FreMa:non}. If $\tilde\phi$ is a function on
$SU_2$ we Fourier transform it to one on the noncommutative space $U(su_2)$ of the spin-model spacetime by
\[ \phi(x)=\int_{SU_2}\extd^3p J(\vec p)\ \tilde\phi(\vec p) \psi_{\vec p}(x),\]
using the noncommutative plane waves 
\[ \psi_{\vec p}(x)=e^{\imath \vec p\cdot\vec x}\]
in \cite{BatMa:non}. Here $x_1,x_2$ and $x_3$ are the generators of  $U(su_2)$ with the commutation  relations \eqref{ncspace} discussed in Section~\ref{spinspace}, and 
 $\extd^3p J(\vec p)$ is the Haar
measure on $SU_2$ in the Lie algebra coordinate system. The orbit spheres in these notations are
\[ C_\m=\{e^{\imath \lambda \vec p\cdot \vec t}\ |\ |\vec p|=\m\}\]
so  $\m=|\vec p|$ defines the sphere, or equivalently 
\[ \P_0=\cos(\m{\lambda/ 2})\]
in our global coordinates $(\P_0,\P_1,\P_2,\P_3)$ of Section~3.2 and  in a patch where $\P_0\ge 0$.
Converting to the corresponding $u$ provides the additional restriction on the spaces $W_{\m s}$ mentioned above as
\[ ({1\over 2}\trace(u)-\cos(\m\lambda/2))\tilde\phi=0.\]

Next, for spin 0 the constraint (\ref{groupconstraint}) on the field $\tilde\phi$ is empty as before and we have to separately impose the $C_\m$ relation as discussed,
\[ \P_0\tilde\phi=\cos(\m{\lambda/ 2})\tilde\phi.\]
Under Fourier transform, multiplication by $\P_0$ becomes $1-\imath{\lambda\over 2}\del_0=\sqrt{1+{\lambda^2\over 4}\Delta}$ in terms of the noncommutative partial derivatives on the noncommutative spacetime. These were introduced in \cite{BatMa:non} but see also \cite{Ma:spo,FreMa:non} (but note the use of $\lambda$ there in the role of $\lambda/2$ in our conventions). All we need to know about the   noncommutative differentials $\del_a$ for the present purposes is that they diagonalise the noncommutative plane waves $\psi_{\vec p}(x)$ with  eigenvalues $\imath\P_a$ . Here $\Delta=\del_a\del^a$ is the noncommutative Laplace operator. So the noncommutative scalar wave equation is
\[ (\Delta+({\sin(\m {\lambda/2})\over\lambda/2})^2)\phi=0.\]
This agrees with \cite{BatMa:non} for a suitable interpretation of the effective mass.

For spin $1/2$ the constraint (\ref{groupconstraint})  is 
\[
e^{\frac \imath  2  \vec p\cdot \vec \sigma}\tilde\phi=e^{-\frac \imath  2  \m\lambda}\tilde \phi.
\] Using our global coordinates, this comes out as
\[ (\P_0+\imath{\lambda\over 2} \vec\P\cdot \vec \sigma)\tilde\phi=e^{-\frac \imath 2 \m\lambda}\tilde\phi\]
Squaring, using the identity $\P_0^2+{\lambda^2\over 4}\vec\P^2=1$ and the constraint equation again to replace $\imath\vec\P\cdot \vec t$, gives the $C_\m$ relations (so these do not need to be imposed separately). Next, using these relations we have
\[ \cos(\m\lambda/2)\tilde\phi+\imath{\lambda\over 2}\vec\P\cdot\vec \sigma\tilde\phi=(\cos(\m\lambda/2)-\imath\sin(\m\lambda/2))\tilde\phi\]
and cancel to obtain 
\[ (\vec\P\cdot\vec  \sigma+{\sin(\m\lambda/2)\over\lambda/2})\tilde\phi=0\]
as the noncommutative Dirac equation in momentum space. This equation squares to  give $\vec\P^2={\sin^2(\m\lambda/2)\over\lambda^2/4}$ which is equivalent to the $C_\m$ relation so this is all we need to impose to obtain the irreducible representation. The equation after Fourier transform becomes
\[ (\imath\vec \del\cdot\vec \sigma-{\sin(\m\lambda/2)\over\lambda/2})\phi=0\]
as the noncommutative Dirac operator for the spin model. This agrees with \cite{BatMa:non}  for our interpretation of the effective mass.

For spin 1, we use the adjoint representation of $SU_2$. The constraint equation (\ref{groupconstraint}) is linear in $\tilde\phi$ so we can use any basis we choose and here we choose the Cartesian one and accordingly work with $\tilde\phi\cdot\vec \sigma$. Then the constraint equation becomes
\[ (\P_0+\imath{\lambda\over 2} \vec\P\cdot\vec \sigma)\tilde\phi\cdot\vec \sigma (\P_0-\imath{\lambda\over 2}\vec\P\cdot\vec \sigma)=e^{-\imath\m{\lambda}}\tilde\phi\cdot\vec \sigma,\]
or
\[ (\P_0^2\tilde\phi\cdot\vec \sigma+\imath{\lambda\over 2} \P_0[\vec\P\cdot\vec \sigma,\tilde\phi\cdot\vec \sigma]+{\lambda^2\over 4}\vec\P\cdot\vec \sigma(\tilde\phi\cdot \vec\P+\imath\tilde \phi\times \vec\P\cdot\vec \sigma))=e^{-\imath\m\lambda}\tilde\phi\cdot\vec \sigma,\]
which comes out as
\[ (1-{\lambda^2\over 2}\vec\P^2)\tilde\phi-\lambda \P_0 \vec\P\times\tilde\phi+
\frac {\lambda^2}{2} (\vec\P\cdot\tilde\phi)\vec\P=e^{-\imath\m\lambda}\tilde\phi.\]
We apply $\vec\P\cdot(\ )$ to both sides and conclude that 
\[ \vec\P\cdot\tilde\phi=0.\]
In spacetime this becomes $\vec\del\cdot\phi=0$ in terms of the noncommutative partial derivatives.
The constraint equation meanwhile reduces to
\[ (2\P_0^2-1-e^{-\imath\m\lambda})\tilde\phi=\lambda  \P_0 \vec\P\times\tilde\phi\]
on replacement of $1-{\lambda^2\over 2}\vec\P^2$. Applying $\vec \P\times$ to this gives
\[ (2\P_0^2-1-e^{-\imath\m\lambda})\vec\P\times\tilde \phi=\frac{4}{\lambda}  \P_0(\P_0^2-1)\tilde\phi\]
on the same replacement. Eliminating $\vec\P\times\tilde\phi$ between these equations gives
an equation for $\P_0$ on $\tilde\phi$ which turns out to be our $C_\m$ relation in the wave operator form. 
Finally, going back to what remained of our constraint equation and replacing $\P_0^2=\cos^2(\m\lambda/2)$ gives 
\[ \vec\P\times\tilde\phi -\imath{\sin(\m\lambda/2)\over\lambda/2}\tilde\phi=0\]
which together with our divergence condition provides the full content of the constraint equation (one may square it to get the $C_\m$ relation once again). Applying the Fourier transform gives
\[ \vec\del\times\phi +{\sin(\m\lambda/2)\over\lambda/2}\phi=0\]
as our spin 1 wave equation, in agreement with \cite{BatMa:non} in the massless case discussed there.

Note that in all these equations, in momentum space the equations in terms of the Lie coordinates $\vec p$ become the same as in the $E_3$ case, since the Lie and global coordinates are related by rescaling with
 $\frac {\sin(\m\lambda/2)}{\m\lambda/2}$, where $\m=|\vec p|$. However, in the noncommutative geometry of $U(su_2)$ it is the $\P_a$ that appear as the natural partial derivatives, see \cite{BatMa:non,Ma:spo}.

\subsection{Representations of $SU_2\times SU_2$}

In this section we show 
that the space \eqref{groupfcts} with a differential
instead of a multiplicative constraint also carries all the irreducible
representations of $SU_2\times SU_2$. This is the semidual model to the $D(SU_2)$  model  of  the preceding section but we shall see that the irreps have a parallel construction.
We denote the generators of the two copies of $su_2$  by
 $J_a^L$ and $J_a^R$;  the Lie brackets are, in our conventions \eqref{brackconv}, 
 \bea
\label{so4brack}
[J_a^L,J_b^L]=\imath\epsilon_{abc} J^L_c, \qquad 
[J_a^R,J_b^R]=\imath\epsilon_{abc} J^R_c, \qquad
[J_a^L,J_b^R]=0.
\eea
The irreps
of  this Lie algebra are well-known to be labelled by two non-negative half-integer spins, which we call  $\n$ and $l$,  and to have dimension $(2\n+1)(2l+1)$. There are two Casimirs
\[(J^R)^2=\sum_{a=1}^3(J_a^R)^2, \quad \mbox{and} \quad (J^L)^2=\sum_{a=1}^3(J_a^L)^2,
\]
which take the following values on the irreps
 \bea
 \label{so4casi}
(J^R)^2=\n(\n+1), \quad (J^L)^2=l(l+1),\quad \n,l\in \frac 1 2(\N\cup{0}).
\eea

We first show that one may realise these operators and  their eigenvalues on the space
\[
W_s=\{\tilde\phi: S^3\rightarrow \CC^{2s+1}\}
\]
of all $\CC^{2s+1}$-valued functions on $S^3$. As before, we let $\rho^s$ be the spin $s$ representation so that
$(\rho^s(t))^2:=\sum_{a=1}^3\rho^s(t_a) \rho^s(t_a)$ has eigenvalue  $s(s+1)$. We define actions of the generators on $W_s$
as 
\bea
\label{so4rep}
J_a^L=\imath\xi^L_a+\rho^s(t_a),\quad J^R_a=\imath\xi^R_a,
\eea
where $\xi^L_a$ and $\xi^R_a$ are the left- and right-generated vector fields associated to the 
generators  $t_a $ of $su_2$ as defined in \eqref{diffdef}.
Squaring, we note that
\[
(J^L)^2=(J^R)^2+2\imath  \xi_a^L \rho^s(t_a)+s(s+1)
\]
 so that \eqref{so4casi}  becomes
 \bea
 \label{casiconst}
(J^R)^2\phi= \n(\n+1)\phi, \quad \phi\in W_s
\eea
 and, with $l=s+\n$,
\bea 
\label{so4constraint}
 ( \imath \xi_a^L\rho^s(t_a)- \n s)\phi=0, \quad \phi\in W_s.
\eea
 This is our `wave equation' in mathematical terms, i.e. we obtain  a representation
on 
\[ W_{ \n s}=\{ \phi:S^3\to \C^{2s+1}\ |\  (\imath \xi_a^L \rho^s(t_a) -\n s)\phi=0\}\]
by imposing this constraint. We still need to impose the condition \eqref{casiconst}
separately in order to obtain an irrep, although this is automatic for spin $1/2$ and $1$ as we shall see shortly. The reason that we then obtain irreps is as follows. We start with the  Peter-Weyl decomposition of $C(SU_2)$ (or rather $L^2$ in a Hilbert space context) in terms of matrix elements of irreps $V_\n$ of $SU_2$. This decomposes the function space into irreducible blocks $V_\n\tens V_\n^*$ where $J^L,J^R$ act on the left and right factors respectively. This is the decomposition provided by the 'wave equation'
\bea \label{so4wave} ((\xi^R)^2+\n(\n+1))\phi=0\eea
on scalar fields (the Laplace-Beltrami equation on $S^3$). Now in our case we have $\C^{2s+1}$-valued fields,
\[ W_s=\C^{2s+1}\tens(\oplus_\n (V_\n\tens V_\n^*))=\oplus_{\n}(\C^{2s+1}\tens V_\n)\tens V^*_\n\]
where $J^R$ acts on $V_\n^*$ as before and $J^L$ acts on $\C^{2s+1}\tens V_\n$. The former is an irrep of $SU_2$ but the latter is not. The constraint \eqref{so4constraint} picks out an irrep of total spin $l=s+\n$ within it. Hence it picks out a block $V_l\tens V_\n^*$ within $W_s$ as isomorphic to our constrained function space $W_{\n s}$ if we also impose (\ref{so4wave}). Hence these are indeed irreps and of the expected size.

It is again interesting to investigate the constraint
\eqref{so4constraint} for low values of $s$. For $s=\frac 1 2 $
we obtain
\bea
\imath\sigma_a\xi^L_a\phi= \n\phi.
\eea
Applying $-\imath\sigma_a\xi^L_a$ to both sides gives
\[
(-(J^R)^2 +i\sigma_a\xi^L_a)\phi= -\imath \n \sigma_a\xi^L_a\phi
\]
or (\ref{so4wave}).

For $s= 1  $
we again  use  the Cartesian representation 
\eqref{cart} to obtain
\bea
\label{spin1}
\epsilon_{abc}\xi^L_a\phi_c=\n\phi_b.
\eea
Acting with $\xi^L_b$ and summing over $c$ gives
\[
-\frac{1}{2}\epsilon_{abc}[\xi^L_a,\xi^L_b]\phi_c=  \n \xi^L_c\phi_c
\Leftrightarrow - \xi^L_c\phi_c = \n \xi^L_c\phi_c.
\]
Since $ \n>0$ we conclude
\[
  \xi^L_c\phi_c=0.
\]
Applying $\epsilon_{deb}\xi^L_d$ to both sides of \eqref{spin1}
now gives
\[
 (J^R)^2\phi_e +\xi^L_d\xi^L_e\phi_d=  \n^2\phi_e.
\]
Now use
\[
\xi^L_d\xi^L_e\phi_d=\xi^L_e\xi^L_d\phi_d +[\xi^L_d,\xi^L_e]
\phi_d = -\n\phi_e
\]
to conclude (\ref{so4wave}) again. 
Thus, like in the euclidean case, only the linear constraint
\eqref{so4constraint} needs to be imposed for $s=\frac 1 2$ and
$s=1$.  

This concludes our wave-equation picture of the representation theory at a mathematical level. In terms of physical variables we can understand the above as follows. We again use a parameter $\lambda$ in parametrising the $SU_2$ where the fields live, but note  that this is now position space and that the value of the parameter in our physical picture is now $\lambda=1/l_c$.  This is the semidual of the model in the preceding section but like that one, it is a (different) deformation of the self-dual $E_3$ model, recovered as $\lambda\to 0$. 
 
Let us note first of all that the actual semidual,  as explained  in Section~3.3,  is $SU_2\rcross SU_2$ by the right adjoint action, which is isomorphic to the above group $SU_2\times SU_2$. Denoting  the generators of the former by $J_a,P_a$ for the two copies respectively, their commutation relations were given in \eqref{jpcoms} and their relations to the generators \eqref{so4brack} are 
\[ P_a=\lambda J^R_a, \quad J_a =J_a^R+J_a^L\;\;\mbox{or}\;\; J_a^L=J_a-\frac {P_a}{\lambda},\quad J^R_a=\frac {P_a} {\lambda}.\]
The physical Casimirs are 
\[ P^2=\lambda^2(J^R)^2,\quad C=\vec P\cdot\vec J-\frac{\lambda} {2}J^2=\frac{\lambda}{2}\left((J^R)^2-(J^L)^2\right).\]
As before, we use the same relations with $p_a$ in place of $P_a$ when we refer to the (noncommutative) momentum space with these as coordinates.

With the definitions \eqref{so4rep},  the action of the angular momentum $J_a$ on the space $W_s$ is
\[ J_a=\imath \Ad_a+\rho^s(t_a),\]
where $\Ad_a=\xi_a^L+\xi_a^R$ is the adjoint action as a vector field on the group in terms of  vector fields for the left and right action \eqref{diffdef} on $SU_2=S^3$. 
This becomes the usual orbital angular momentum on $\RR^3$
in the limit $\lambda\to 0$. The action of $P_a$ is
\[ P_a=\imath \lambda \xi^R_a\]
and the associated Casimir is the Laplace-Beltrami operator on $S^3$. Its eigenvalues (the squared mass of the particle) are,
according to \eqref{so4wave}, given by
\bea
\label{psquared}
P^2\phi =  \lambda^2 \n( \n+1)\phi,
\eea
so essentially $\m= \lambda \n$ is the mass of the particle. 

Next, a short computation gives
\[ C=-\imath\lambda\xi_a^L \rho^s(t)_a-\frac{\lambda}{2} s(s+1).\]
In line with what we have done before, we therefore impose a suitable value of this as a further  `wave operator' to 
 obtain representations of $SU_2\times SU_2$ on the spaces 
\[ W_{ \n s}=\{\phi:S^3\to \C^{2s+1}\ |\ (\vec P\cdot \vec J-\frac{\lambda}{2} J^2+ \lambda \n s+\frac \lambda 2 s(s+1))\phi=0\},\]
which are irreps at least for spin $1/2$ and spin $1$. For higher spin we need to impose \eqref{so4wave} as well.
Taking the limit $\lambda \rightarrow 0$ while keeping the mass $\m  = \lambda \n$ fixed reproduces the constraint
\eqref{posconstraint}
in euclidean space, as required.

Note that these computations are done in position space. In terms of our previous exposition, we have gone from noncommutative momentum space (functions of the $p_a$) to position space (functions on $SU_2$) again by means of the quantum group Fourier transform, this time read the other way. The only fact we need to know is that left multiplication by $p_a$ becomes the vector field $-\imath \xi_a^L$ while right-multiplication by $p_a$ becomes the vector field $\imath \xi^R_a$. If one wants to do things in the noncommutative momentum space then the constraint (\ref{so4constraint}) appears as
\[( \rho^s(t_a)p_a+\m s)\tilde\phi(\vec p)=0.\] 
We are distinguishing here between the generators $P_a$ of the isometry group and the noncommutative coordinates $p_a$ on momentum space. They are both copies of the  scaled $su_2$ Lie algebra relations as stated for $P_a$ above. 

Note that our `orbits' or conjugacy classes in momentum space still exist as before, but now as `fuzzy spheres' of radius $\m=\lambda \n$ in this momentum space instead of usual spheres as for the $E_3$ model.  It is known how to construct monopole sections
in this context (as projective modules) but we are not aware of a full analogue of the Hopf fibration itself, hence the `upstairs' point of view with field $\psi$ as in \eqref{doublerep}
 requires further elaboration using methods of noncommutative geometry. The downstairs picture of the monopole sections is defined for $s=1/2$ by 
projections
\[ e({\vec P})=\frac {\n+1 }{ 2(\n+\frac 1 2)}+\frac {t_aP_a}{\lambda(\n+\frac 1 2)}.\]
One can check that $e^2=e$ using the $P_a$ commutation relations \eqref{jpcoms}
 and  the constraint  \eqref{psquared}. As $\lambda\to 0$
and $\n \rightarrow \infty$  with  $\m=\lambda \n$  fixed we see that  we recover the standard monopole projector  given in \eqref{epspinhalf}. 

\section{Discusson}

We have seen that the `particle content' in the $E_3$ flat spacetime model can be deformed in two ways, one with the mass $\m$ `compressed' by  the  sine function  as momentum space is compactified to $SU_2$  but otherwise similar (the spin model) and the other with mass $\m$ discretised in units of $\lambda$ due to a fuzzy sphere in momentum space (the $SU_2\times SU_2$ model). Thus although the physical parameters for the irreps in the two models are very different the actual constructions of the irreps are similar and in some sense the physical states `correspond' through their common limit (i.e. with arbitrary accuracy as the relevant $\lambda\to 0$) even though they are different.  This is the `remnant' of the self-duality in the degenerate cases  that we have looked at (the upper part of Figure~1). 

This picture also applies elsewhere in Figure~1 and can, in principle, be developed entirely analogously. Thus the the $SL_2(\C)$ model of Section~3.4 is similar in principle to the  $SU_2\times SU_2$ model of Section~4.3 while its semidualisation is the bicrossproduct model. Its representation theory, as a semidirect product algebra, is readily developed in the same manner as for the quantum double in Section~4.2. The difference is that the adjoint action is replaced by a non-linear action deforming it as we have explained in Section~3.5. In both cases we have complications due to the non-compactness. The `quantum gravity with cosmological constant' case of Section~3.6 can similarly be developed -- with a lot more effort -- as a $q$-deformation of Section~4.2. Here again we see that the irreps on the one hand are
those of quantum $SL_2(\C)$ and on the other hand in the semidual model, they are irreps of quantum $SU_2\times SU_2$ -- described by the same parameters as in the non-$q$-deformed case and with the same features of continuous and discrete parameters being `matched' in a limiting sense. How this proceeds given that the signatures (expressed in the $*$-structures) are very different remains to be seen. Roughly speaking, we expect that the algebraic equivalence of categories ignoring the $*$-structures explained in Section~3.6 is complemented by  two different `cross sections' consisting of the unitary  irreps  in each model,
and that these are slices are in some sense `transverse'. 

We can gain some insight again from the simplest $E_3$ case. Thus here on the one hand we have irreps of $E_3$ constructed as monopole sections over spheres and a dual model in which the irreps are constructed by wave equations in $\R^3$. In a fixed point of view these are respectively momentum and position space treatments but from the point of view in which each theory is considered the primary one, they are both (say) position space representations. Thus we consider functions $\phi(\vec x)$ with values in $\C^{2s+1}$ and consider both our possible constraints as two different physical models on this position space $\R^3$. One is related to the operator $\rho^s(t_a)\del_a$ and the other to the operator $\rho^s(t_a)  x_a$. It is interesting to note that for $s=1/2$ these two are closely related to the Riemannian geometry of the sphere. Thus,
 \[ [\vec t\cdot \vec  x, \vec t\cdot\nabla]=\imath t\cdot(\vec x\times\nabla)-\frac 3 4,\]
 using elementary properties of the Pauli matrices. Now the expression on the right is essentially a massive Dirac operator on a sphere $S^2$ with its standard Riemannian metric. (It commutes with $x^2$ and hence defines an operator on $\C^2$-valued functions on the sphere.) Thus Riemannian geometry arises here out of the interaction of the system and the dual system. Also, we see that our two operators form some kind of `Heisenberg pair' with the curved Dirac operator in the role of Planck's constant.  In this sense, our two methods of extracting irreps of $E_3$ are `transverse' and describe different physics if one views both in position space, in the sense that one cannot simultaneously restrict to both: restricting to an irrep in one point of view should typically have inner products with all the irreps in the other point of view. We expect that this is part of the  story for the full quantum gravity case.  

\appendix

\section{Vector fields and forms on Lie groups}
\label{lieapp}
Here we collect some facts about forms and vector fields on  
an $n$-dimensional Lie group $G$, which are used in the main text.
In order  to simplify notation we assume $G$ to be a matrix group.
We write $\g$ for the Lie algebra of $G$, and work with
generators for which the structure constants are purely imaginary.
With the notation $t_a, a=1,\ldots, n$, for the generators  the  Lie brackets take the form
\bea
\label{brackconv}
[t_a,t_b]=\imath f^c_{ab} t_c, 
\eea 
where the $f^c_{ab}$ are real, and  we use the convention that repeated indices are summed over. 
It follows that  the structure constants are $f^c_{ab}$  in terms of the "real" generators
 $-\imath t_a$; the reader may find it useful to read some of the geometrical formulae in this paper 
 in terms of these generators.
Associated to the generators $t_a$ 
we have the left-generated vector fields $\xi_a^L$
and the right-generated vector fields $\xi_a^R$,  defined via
\bea
\label{diffdef}
\xi_a^L f(g) =\frac{d}{ds}|_{s=0}f(e^{\imath s t_a}g), \qquad 
\xi_a^R f(g) =\frac{d}{ds}|_{s=0}f(ge^{-\imath s t_a}).
\eea
They close under the Lie bracket of vector fields,
and give  two commuting copies of 
$\g$:
\bea
[\xi_a^L,\xi_b^L]=f^c_{ab} \xi^L_c, \qquad 
[\xi_a^R,\xi_b^R]=f^c_{ab} \xi^R_c, \qquad
[\xi_a^L,\xi_b^R]=0 \nonumber 
\eea 
Using the matrix structure of $G$ we can identify  $T_gG$
with matrices of the form $g\xi$, where $\xi \in \g$,
or with  matrices of the form $\xi g $.
Then we can also write
\bea
\label{matrixdef}
\xi_a^L(g)=\imath t_ag, \qquad \xi_a^R(g)=-\imath gt_a.
\eea
Using either of the definitions \eqref{diffdef} and \eqref{matrixdef}
it is easy to see that the left-generated vector fields
are invariant under the right-action  $R_h:g\mapsto gh$ of $G$
on itself (and hence
on $TG$) and that the 
 right-generated vector fields
are invariant under the  left-action $L_h:g\mapsto hg$  of $G$
on itself.
We have the following relation between left- and right-generated
vector fields: 
\bea
(L_g R_{g^{-1}})'(\xi_a^L(g)=-\xi_a^R(g). \nonumber
\eea
With the abbreviation
\[
\Ad(g)(t_a)=gt_ag^{-1}=R^b_{\;\;a}(g) t_b
\]
it follows that  
\[
\xi_a^R(g)=-R^b_{\;\;a}(g)\xi_b^L(g).
\]

There is as basis of one-forms dual to the above vector fields
which can be obtained by expanding the 
Maurer-Cartan form
\[
\theta =g^{-1} \extd g.
\]
The Maurer-Cartan form is Lie-algebra valued, and manifestly
left-invariant. 
Expanding in the Lie algebra basis $t_a$, $a=1,\ldots, n$,
we obtain a basis $\sigma^{R,a}$ of left-invariant one-forms
\bea
\label{expandr}
g^{-1} \extd g= -\imath t_a\sigma^{R,a}.
\eea
The one forms $\sigma^{R,a}$ are dual to the left-invariant
(and right-generated) vector fields $\xi_a^R$:
\[
\sigma^{R,a}(\xi_b^R)=\delta^a_{\;\;b}.
\]
We obtain right-invariant one-forms $\sigma^L_a$ by expanding
\bea
\label{expandl}
-g \extd(g^{-1})=\extd g g^{-1}=\imath t_a\sigma^{L,a}
\eea
with the duality relation
\[
\sigma^{L,a}(\xi_b^L)=\delta^a_{\;\;b}.
\]
Comparing \eqref{expandr} with \eqref{expandl}
we have the relation
\[
\sigma^{L,a}=-R^a_{\;\;b}\sigma^{R,b}.
\]
Since the Maurer-Cartan form satisfies
\[
\extd \theta + \theta \wedge \theta =0
\]
we deduce
\[
\extd \sigma^{R,a}=-\frac 1 2 f_{bc}^a\sigma^{R,b}\wedge\sigma^{R,c}
\]
and by a similar argument
\[
\extd \sigma^{L,a}=-\frac 1 2 f_{bc}^a\sigma^{L,b}\wedge\sigma^{L,c}.
\]

We note that every compact Lie group  has a  bi-invariant  Riemannian
metric.
In terms of the one-forms introduced above  it can be written 
\bea
\label{bimetric}
\extd s^2=\kappa_{ab}\sigma^{R,a}\sigma^{R,b}=\kappa_{ab}\sigma^{L,a}\sigma^{L,b}
\eea
where $\kappa$ is the Killing form on the Lie algebra i.e.
\[
\kappa_{ab}=-\tr(\ad(t_a)\ad(t_b)).
\]
The Laplace operator associated to this metric can be written  in terms of the 
inverse metric $\kappa^{ab}$  and either the left- or right-generated vector fields as 
\bea
\label{leftisright}
 \kappa^{ab} \xi^R_a \xi^R_b= \kappa^{ab} 
\xi^L_a \xi^L_b.
\eea

Finally, although the tangent bundle of any Lie group is isomorphic to the trivial bundle $G\times \g$, this is not canonical in the sense that we can use either the left- or the right-translations
to trivialise the bundle. In the 
left-trivialisation, $g \xi\in T_gG$ is identified with $\xi \in
\g$.
In the right-trivialisation  $\xi g \in T_gG$ is identified with $\xi \in
\g$. Both left- and right-translation can also 
be used to define a  connection  on $TG$. Both the connections are
flat. In the left-trivialisation,
the connection defined by the left-translation has the covariant
derivative $D_L=\extd$. The right-translation has the covariant
derivative $D_R=\extd+g^{-1}\extd g$. Note that $D^2_R=0$, as required for 
flatness.
The Levi-Civita connection (unique torsion free connection which 
preserves the Killing metric \eqref{bimetric})  turns out to
be the average of the connection for the leff- and right-translation. 
In the left-trivialisation the Levi-Civita
connection one-form is therefore 
$A_{LC}+\frac 1 2 g^{-1}\extd g$, leading to the covariant
derivative $D_{LC}=\extd+\frac 1 2 g^{-1}\extd g$. The Levi-Civita connection 
is not flat. Its curvature is 
\[
F_{LC}=\extd(\frac 1 2 g^{-1}\extd g) +\frac 1 4 
g^{-1}\extd g\wedge g^{-1}\extd g = -\frac 1 4 g^{-1}\extd g\wedge g^{-1}\extd g.
\]

\end{document}